\documentclass[aps,prd,onecolumn,groupedaddress,showpacs,nofootinbib,amssymb
]{revtex4}
\usepackage[dvips]{graphicx}
\usepackage{amssymb}
\usepackage{amsmath}
\usepackage{graphicx}
\usepackage{amsfonts}
\usepackage{bm}

\begin{document}

\title{Singular Inflationary Universe from $F(R)$ Gravity}
\author{
S.~D.~Odintsov,$^{1,2}$\,\thanks{odintsov@ieec.uab.es}
V.~K.~Oikonomou,$^{3,4}$\,\thanks{v.k.oikonomou1979@gmail.com}}
\affiliation{ $^{1)}$Institut de Ciencies de lEspai (IEEC-CSIC),
Campus UAB, Carrer de Can Magrans, s/n\\
08193 Cerdanyola del Valles, Barcelona, Spain\\
$^{2)}$ ICREA, Passeig LluA­s Companys, 23,
08010 Barcelona, Spain\\
$^{3)}$ Tomsk State Pedagogical University, 634061 Tomsk, Russia\\
$^{4)}$ Lab. Theor. Cosmology, Tomsk State University of Control Systems
and Radioelectronics, 634050 Tomsk, Russia (TUSUR)\\
}

\begin{abstract}
Unlike crushing singularities, the so-called Type IV finite-time singularity offers the
possibility that the Universe passes smoothly through it, without any
catastrophic effects. Then the question is if the effects of a Type IV singularity can be
detected in the process of cosmic evolution. In this paper we address this
question in the context of $F(R)$ gravity. As we demonstrate, the effects
of a Type IV singularity appear in the Hubble flow parameters, which
determine the dynamical evolution of the cosmological system. So we study
various inflation models incorporating a Type IV singularity, with the
singularity occurring at the end of inflation. Particularly we study a toy
model and a singular version of the $R^2$ gravity Hubble rate. As we evince, some of
the Hubble flow parameters become singular at the singularity, an effect
which indicates that at that point a dynamical instability occurs. This
dynamical instability eventually indicates the graceful exit from inflation.
We demonstrate that the toy model has an unstable de Sitter point at the singularity, so
indeed graceful exit could be triggered. In the case of the singular inflation
model, graceful exit proceeds in the standard way. In addition, we
investigate how the form of the $F(R)$ gravity affects the singularity
structure of the Hubble flow parameters.
 In the case of the singular inflation model, we found various scenarios for
singular evolution, most of which are compatible with observations, and only
one leads to severe instabilities. In addition, in one of these scenarios, the presence of the Type IV singularity, slightly modifies the spectral index of primordial curvature perturbations. We
also compare the ordinary Starobinsky with the singular inflation model, and we
point out the qualitative and quantitative differences. Finally, we study
the late-time dynamics of the toy model and of the singular inflation model and
we demonstrate that the unification of early and late-time acceleration can
be achieved. We also show that it is possible to achieve late-time
acceleration similar to the $\Lambda$-Cold Dark Matter model.
\end{abstract}


\maketitle



\def\pp{{\, \mid \hskip -1.5mm =}}
\def\cL{\mathcal{L}}
\def\be{\begin{equation}}
\def\ee{\end{equation}}
\def\bea{\begin{eqnarray}}
\def\eea{\end{eqnarray}}
\def\tr{\mathrm{tr}\, }
\def\nn{\nonumber \\}
\def\e{\mathrm{e}}

\section{Introduction}

Singularities in physical theories always indicate that the theory needs
some modification at the scales that the singularity occurs, since the
singularity itself reveals an incompleteness of the theoretical description.
In General Relativity, there are two types of singularities, the timelike and
spacelike singularities, both described by the Hawking-Penrose theorems
\cite{hawkingpenrose}. Spacelike singularities occur in compact objects like
black holes \cite{Virbhadra:2002ju}, while timelike singularities are in
some way some events that occur globally on a spacelike hypersurface. The
timelike singularities are known in cosmology as finite time singularities
and where firstly extensively classified in \cite{Nojiri:2005sx}. Among
these, there are crushing type singularities, like the Big Rip \cite{ref5},
but also milder singularities, like the Type IV singularity
\cite{Nojiri:2005sx,Barrow:2015ora,noo1,noo2,noo6}. In the
same class of milder singularities belong the sudden singularities, firstly
studied in \cite{Barrow:2004hk,Barrow:2004xh}, and further developed in
\cite{barrow}. The difference between crushing types singularities and
milder singularities is that, in the crushing type singularities geodesics
incompleteness occurs and therefore these singularities are catastrophic
events, since some or all the physical quantities defined on a spacelike
hypersurface at the moment that the singularity occurs, strongly diverge.
However, in the case of milder singularities, like the Type IV, the physical
quantities do not diverge and therefore the passage of the Universe through
such a singularity is smooth. However, the existence of the singularity can
be responsible for a number of interesting phenomenological consequences, as
was demonstrated in \cite{noo1,noo2,noo6}. Particularly, as
was shown in \cite{noo2}, the existence of a Type IV singularity in a
scalar-tensor cosmological framework, generates an instability in the second
Hubble slow-roll parameter \cite{barrowslowroll}, a fact that can indicate a
possible mechanism for the graceful exit from inflation for that
scalar-tensor model. For reviews and useful studies on inflation, the reader
is referred to \cite{inflation} and in addition, for alternative ideas with
regards to the graceful exit one, see \cite{lehners}. In the model studied
in \cite{noo2}, it was assumed that the slow-roll condition for the
canonical scalar field did not hold true, and the exit from inflation could
occur owing to the instability of the second Hubble slow-roll index
$\eta_H$.

In this paper we shall investigate the phenomenological implications of a
Type IV singularity on the inflationary dynamics of vacuum $F(R)$ gravity.
Particularly, we shall study a toy model inflationary solution containing a
mild singularity at a certain time instance, and we also find the $F(R)$
gravity (for review, see \cite{reviews1}) that can generate such an
inflationary solution. In addition to this toy model, we also study a
singular analog of $R^2$ inflation \cite{starobinsky}. In addition, we
shall calculate the corresponding Hubble flow (also called slow-roll)
parameters \cite{encyclopedia,noh}, and we will investigate what are the
effects of the Type IV singularity on the Hubble flow parameters for the
aforementioned models. The Hubble flow parameters govern the inflationary
dynamics, so if the dynamics are affected directly by the presence of the
Type IV singularity, this would be a clear indication that the dynamics is
interrupted or modified at the singularity point, and therefore the final
attractor solution can be changed at exactly the singularity point. As we
shall demonstrate, for the inflationary models we shall study, the Hubble
flow parameters are small for small cosmic times, but some of the parameters
blow up at the singularity point. Therefore, the dynamical evolution at that
point is abruptly interrupted and we interpret this instability as an
indication that inflation ends at the moment that the singularity occurs.
Therefore, the Type IV singularity acts as an indicator of another possible
mechanism for graceful exit from inflation in $F(R)$ gravity, to be added in
the well established mechanisms for graceful exit in the Jordan frame
\cite{starobinsky,sergeitraceanomaly}. We need to note that the form of the
$F(R)$ gravity can potentially modify the singularity structure of the
Hubble flow parameters, and in order to reveal this we adopt two different
approaches, by studying the $F(R)$ gravity near the Type IV singularity and
also by assuming a more general form for the vacuum $F(R)$ gravity. An
important result of our analysis is that the inflationary picture is not
affected by the Type IV singularity, when physical quantities are
considered, but the dynamics of inflation are affected strongly and only at
the singularity point. In addition, and in support to this, even quantities
constructed from physical quantities, like the comoving Hubble radius, are
totally unaffected from the Type IV singularity. We are therefore have at
hand a quite physically appealing situation for which the physics on the
three dimensional spacelike hypersurface corresponding to the singularity
time, can be defined in a singularity free way, and the implications of the
singularity can be found in auxiliary parameters that determine the
dynamics. In addition, the slow-roll condition, a necessary condition in
order a sufficient number of $e$-foldings is achieved, also holds true in
our case, and in fact by choosing appropriately the time at which the
singularity occurs, we can produce the desirable number of $e$-foldings,
since inflation can end at the singular point. The most important outcomes
are obtained for the singular inflation model, in which case we demonstrate that
the singular model can be compatible with current observational data in most
cases, and only in one case strong instabilities occur, which make this
situation peculiar. Finally, the models of inflation we shall study, also provide
an interesting late-time behavior, which we analyze by investigating the
corresponding Equation of State (EoS). Particularly, as we evince, it
possible that the singular cosmological models we studied, offer a
consistent theoretical framework, in the context of which, early-time and
late-time acceleration are unified \cite{sergnoj,capp}. For the singular
inflation model, this is a particularly appealing scenario, since it is possible
to describe early-time acceleration consistent with current observational
data, but also with the same model, late-time acceleration can also be
consistently described.

This paper is organized as follows: In section II we provide in brief all
the essential information for the finite time singularities and also present
all the conventions we shall assume that hold true throughout the paper. In
section III, we present the toy model Type IV singular inflationary solution
and in section IV we investigate which $F(R)$ gravity can successfully
describe the toy model inflationary solution with emphasis being given for
times near the Type IV singularity. In addition, in section V we analyze in
detail the inflationary picture of the toy model and we calculate in detail
the Hubble flow parameters, for the corresponding solution of the $F(R)$
gravity we found. Moreover, we comment on the singularity structure of
the Hubble flow parameters for the case that the $F(R)$ gravity has a more
general form. Moreover, we analyze the graceful exit mechanism that the
infinite instabilities of the Hubble flow parameters seem to impose.
Additionally, we demonstrate that graceful exit can be achieved owing to the
existence of an unstable de Sitter point at the singularity. In section VI
we analyze in detail a singular version of the Starobinsky $R^2$ inflation model in the Jordan frame, and we present all the different evolution
scenarios. We also discuss the implications of the different scenarios, both
on observations and also on the dynamics of the cosmological system. In
section VII we study the late-time dynamics of all the cosmological models
we analyzed in the previous sections and we demonstrate that all the models
succeed to describe the unification of early with late-time acceleration. The
conclusions along with a discussion follow in the end of the paper.

\section{Finite Time Singularities Essentials and Conventions}

Before we discuss the singular inflation models we shall use in this paper,
it is worth to recall some basic information with regards to finite time
singularities. The detailed classification of finite time cosmological
singularities was done in Ref.~\cite{Nojiri:2005sx}, and the classification
uses three physical quantities, the effective energy density, the effective
pressure, the scale factor and also the Hubble rate and it's higher
derivatives. It is worth noting that all the finite time singularities are
timelike singularities. The finite time singularities are classified in the
following way \cite{Nojiri:2005sx},
\begin{itemize}
\item Type I (``Big Rip Singularity''): It is the most ``severe''
singularity from a phenomenological point of view, and it occurs when the
cosmic time $t$ approaches a time instance $t_s$, that is, $t\rightarrow
t_s$,  the scale factor $a$, the effective energy
density $\rho_{\mathrm{eff}}$ and also the effective pressure
$p_\mathrm{eff}$ diverge, that is, $a \to \infty$,
$\rho_\mathrm{eff} \to \infty$, and $\left|p_\mathrm{eff}\right| \to
\infty$ respectively. For more details on this singularity, the reader is
referred in Refs.~\cite{ref5}.
\item Type II (``Sudden Singularity''): In this case, as $t \to t_s$, both
the scale factor and the effective energy density are finite, that is, $a
\to a_s$, $\rho_{\mathrm{eff}}\to \rho_s$, but the effective pressure
diverges, $\left|p_\mathrm{eff}\right| \to \infty$. For more information on
this type of singularity, consult \cite{Barrow:2004hk,Barrow:2004xh,barrow}.
\item Type III: For this finite time singularity, as the cosmic time
approaches the time instance $t_s$, both the effective energy density and
the effective pressure diverge, that is $\rho_\mathrm{eff} \to \infty$ and
$\left|p_\mathrm{eff}\right| \to \infty$, but the scale factor remains
finite, $a \to a_s$.
\item Type IV: The most mild from a phenomenological point of view, since in
this case, the Universe can smoothly pass through it, with all the physical
quantities that can be defined on the three dimensional spacelike
hypersurface $t=t_s$, remaining finite. Particularly, the scale factor, the
energy density and the pressure remain finite, that is,  $a \to a_s$,
$\rho_\mathrm{eff} \to \rho_s$,
$\left|p_\mathrm{eff}\right| \to p_s$. In addition, the Hubble rate and it's
first derivative also remain finite, but the higher derivatives, or some
higher derivative, diverge. For some phenomenological consequences of this singularity, see \cite{noo1,noo2,noo6}.
\end{itemize}
As we already mentioned, the focus in this paper will be on the Type IV
singularity, which we study in the Jordan frame. Before proceeding we need
to specify the geometric background we shall assume. We will consider a
Jordan frame $F(R)$ gravity in the metric formalism, with the background
metric being the flat Friedmann-Robertson-Walker (FRW),
\be
\label{metricfrw} ds^2 = - dt^2 + a(t)^2 \sum_{i=1,2,3}
\left(dx^i\right)^2\, ,
\ee
where $a(t)$ denotes the scale factor. Moreover, we shall make use of a
torsion-less, symmetric, and
metric compatible affine connection, the Levi-Civita connection.

\section{$F(R)$ Gravity Description Near the Type IV Singularity: A Singular
Toy Model}

Having described the general features of finite time singularities, we now
proceed to the description of the toy model we shall use for warm-up, which
is an inflationary solution containing a Type IV finite time singularity. In
this section we present the toy inflationary solution, and we also
investigate which vacuum $F(R)$ gravity can generate such a cosmological
evolution, emphasizing to cosmic times near the Type IV singularity.

The main feature of the toy inflationary solution is that it produces an
inflationary era, so for a long time, the toy inflationary solution should
be a de Sitter solution. Also, we choose the Type IV singularity to occur at
the end of the inflationary era. To state this more correctly, the Type IV
singularity indicates when the inflationary era ends. In the following
sections we shall provide sufficient indications to support this argument.

\begin{figure}[h] \centering
\includegraphics[width=15pc]{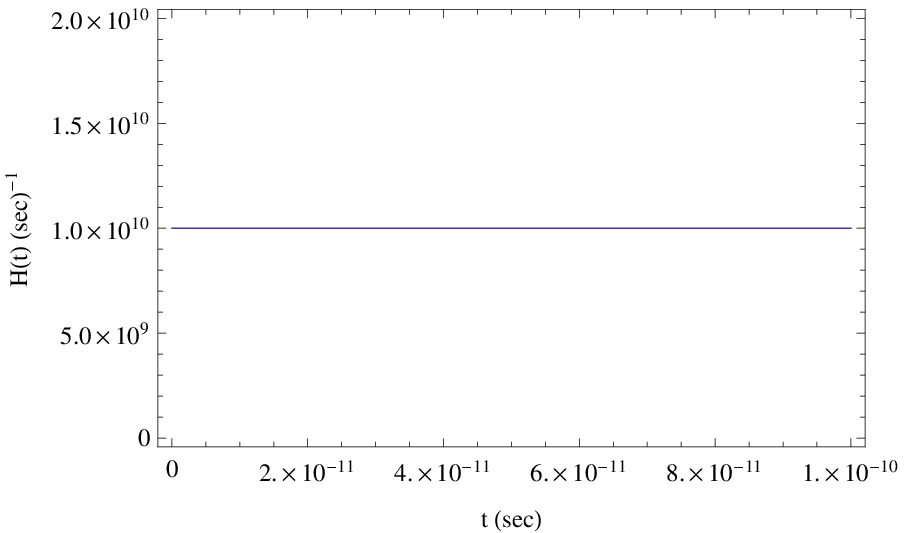}
\includegraphics[width=15pc]{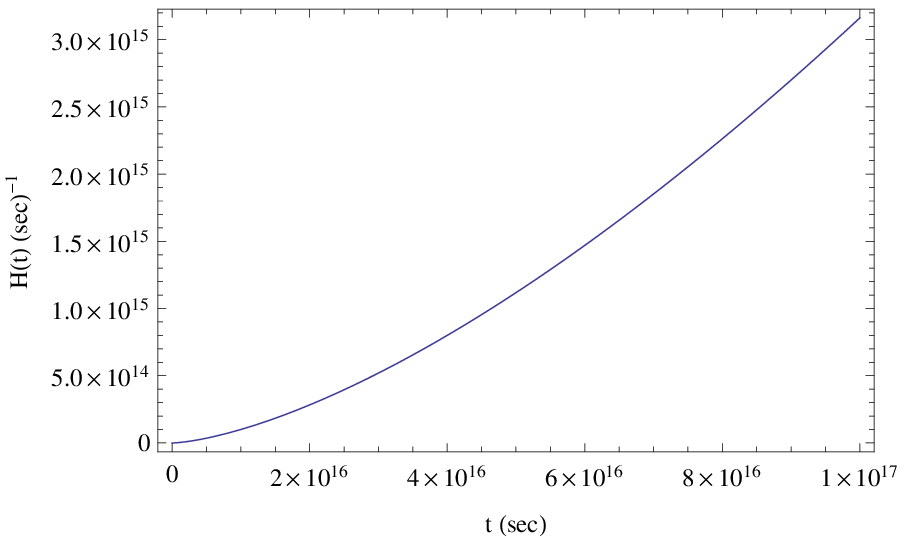}
\caption{The Hubble rate $H(t)$ as a function of the cosmic time $t$, for
$t_s=10^{-35}$sec, $\alpha=3/2$, $f_0=10^{-10}(\mathrm{sec})^{-\alpha-1}$,
and $c_0=10^{10}(\mathrm{sec})^{-1}$}
\label{plot1}
\end{figure}

The toy inflationary solution which we shall describe, is described by the
following Hubble rate,
\begin{equation}
\label{hublawsing}
H(t)=c_0+f_0\left( t-t_s \right)^{\alpha}\, ,
\end{equation}
with the assumption that $c_0\gg f_0$ and also for the cosmic times near the
inflationary era, it holds true that $c_0\gg f_0\left( t-t_s
\right)^{\alpha}$, for $\alpha>0$. So in effect, near the time instance
$t\simeq t_s$, the cosmological evolution is a nearly de Sitter.
Also, the Type IV singularity occurs at $t=t_s$, as it can be seen from Eq.
(\ref{hublawsing}). Particularly, the singularity structure of the
cosmological evolution (\ref{hublawsing}), is determined from the values of
the parameter $\alpha$, and for various values of $\alpha$ it is determined
as follows,
\begin{itemize}\label{lista}
\item $\alpha<-1$ corresponds to the Type I singularity.
\item $-1<\alpha<0$ corresponds to Type III singularity.
\item $0<\alpha<1$ corresponds to Type II singularity.
\item $\alpha>1$ corresponds to Type IV singularity.
\end{itemize}
So in order to have a Type IV singularity we must assume that $\alpha>1$,
and we adopt this constraint for the parameter $\alpha$ in the rest of this
paper. For $\alpha>1$, the cosmological evolution near the Type IV
singularity is a nearly de Sitter evolution. Indeed, since $c_0\gg f_0$, the
term $\sim f_0\left( t-t_s \right)^{\alpha}$ is negligible at early times,
but it can easily be seen that it dominates the evolution at late times. In
Fig. \ref{plot1}, we plot the Hubble rate (\ref{hublawsing}) at early times,
by choosing $t_s=10^{-35}$sec. Also we chose $c_0=10^{10}$sec and
$f_0=10^{-10}$$\mathrm{sec}^{-1-\alpha}$. Note that we measure time in
seconds, so the Hubble rate is measured in $\mathrm{sec}^{-1}$ and also the
present time corresponds to $t_p\sim 10^{17}$sec.
As it can be seen from Fig. (\ref{plot1}), the evolution is governed by
$c_0$ at early times and for a sufficient period of time after $t=t_s$, and
the evolution is governed by the term $\sim f_0\left( t-t_s
\right)^{\alpha}$ only at late times $\sim t_p$. In principle the variables
can be adjusted to achieve a more phenomenological acceptable cosmology, but
we simply demonstrated the behavior of $H(t)$ as a function of time for
illustrative reasons, and in order to stress the fact that at early times,
the term $c_0$ governs the evolution, and the second term $\sim f_0\left(
t-t_s \right)^{\alpha}$ governs the late-time evolution. Also it is
important to note that the singularity essentially plays no particular role
when one considers the Hubble rate and other observable quantities at early
times. But as we shall demonstrate, it plays a crucial role in the dynamical
evolution. In the FRW background of Eq. (\ref{metricfrw}), the Ricci scalar
reads,
\begin{equation}
\label{ricciscal}
R=6(2H^2+\dot{H})\, ,
\end{equation}
so for the Hubble rate of Eq. (\ref{hublawsing}), the Ricci scalar reads,
\begin{equation}
\label{desitterricci}
R=12 c_0^2+24 c_0 f_0 (t-t_s)^{\alpha }+12 f_0^2 (t-t_s)^{2 \alpha }+6 f_0
(t-t_s)^{-1+\alpha } \alpha \, ,
\end{equation}
and consequently near the Type IV singularity, the Ricci scalar is $R\simeq
12 c_0^2$.

\subsection{$F(R)$ Gravity Description}

We now investigate which vacuum $F(R)$ gravity can generate the cosmological
evolution described by the Hubble rate (\ref{hublawsing}). We shall use the
reconstruction method used in Refs.
\cite{Nojiri:2006gh,Capozziello:2006dj,sergbam08} and we are
mainly interested in the description near the Type IV singularity.

The action of a vacuum Jordan frame $F(R)$ gravity is equal to,
\begin{equation}
\label{action1dse}
\mathcal{S}=\frac{1}{2\kappa^2}\int \mathrm{d}^4x\sqrt{-g}F(R)\, ,
\end{equation}
and by adopting the metric formalism, we vary the action of Eq.
(\ref{action1dse}) with respect to the metric $g_{\mu,\nu}$, so we obtain
the following Friedmann equation,
\begin{equation}
\label{frwf1}
 -18\left [ 4H(t)^2\dot{H}(t)+H(t)\ddot{H}(t)\right ]F''(R)+3\left
[H^2(t)+\dot{H}(t)
\right ]F'(R)-
\frac{F(R)}{2}=0\, .
\end{equation}
The reconstruction method we shall adopt, makes use of an auxiliary scalar
field $\phi$, so the $F(R)$ gravity of Eq. (\ref{action1dse}) can be written
in the following equivalent form,
\begin{equation}
\label{neweqn123}
S=\int \mathrm{d}^4x\sqrt{-g}\left [ P(\phi )R+Q(\phi ) \right ]\, .
\end{equation}
Note that the auxiliary field has no kinetic form so it is a non-dynamical
degree of freedom. The reconstruction method we employ is based on finding
the analytic dependence of the functions $P(\phi )$ and $Q(\phi )$ on the
Ricci scalar $R$, which can be done if we find the function $\phi (R)$.  In
order to find the latter, we vary the action of Eq. (\ref{neweqn123}) with
respect to $\phi$, so we end up to the following equation,
\begin{equation}
\label{auxiliaryeqns}
P'(\phi )R+Q'(\phi )=0\, ,
\end{equation}
where the prime in this case indicates the derivative of the corresponding
function with respect to the auxiliary scalar field $\phi$. Then by solving
the algebraic equation (\ref{auxiliaryeqns}) as a function of $\phi$, we
easily obtain the function $\phi (R)$. Correspondingly, by substituting this
to Eq. (\ref{neweqn123}) we can obtain the $F(R)$ gravity, which is of the
following form,
\begin{equation}
\label{r1}
F(\phi( R))= P (\phi (R))R+Q (\phi (R))\, .
\end{equation}
Essentially, finding the analytic form of the functions $P(\phi)$ and
$Q(\phi)$, is the aim of the reconstruction method. These can be found by
varying the action of Eq. (\ref{neweqn123}) with respect to the metric
tensor $g_{\mu \nu}$, and the resulting expression is,
\begin{align}
\label{r2}
 & -6H^2P(\phi (t))-Q(\phi (t) )-6H\frac{\mathrm{d}P\left (\phi (t)\right
)}{\mathrm{d}t}=0\, , \nn
& \left ( 4\dot{H}+6H^2 \right ) P(\phi (t))+Q(\phi (t)
)+2\frac{\mathrm{d}^2P(\phi
(t))}{\mathrm
{d}t^2}+\frac{\mathrm{d}P(\phi (t))}{\mathrm{d}t}=0\, .
\end{align}
By eliminating the function $Q(\phi (t))$ from Eq. (\ref{r2}), we obtain,
\begin{equation}
\label{r3}
2\frac{\mathrm{d}^2P(\phi (t))}{\mathrm {d}t^2}-2H(t)\frac{\mathrm{d}P(\phi
(t))}{\mathrm{d}t}+4\dot{H}P(\phi (t))=0\, .
\end{equation}
Hence, for a given cosmological evolution with Hubble rate $H(t)$, by
solving the differential equation (\ref{r3}), we can have the analytic form
of the function $P(\phi)$ at hand, and from this we can easily find $Q(t)$,
by using the first relation of Eq. (\ref{r2}). Note that, since the action
of the $F(R)$ gravity (\ref{action1dse}) with the action
(\ref{neweqn123}) are mathematically equivalent, the auxiliary scalar field can be identified with the
cosmic time $t$, that is $\phi=t$. For more details on this, see for example
the Appendix of Ref.~\cite{Nojiri:2006gh}.

Having described the general reconstruction method, let us now apply it for
the cosmology described by the Hubble law of Eq. (\ref{hublawsing}),
emphasizing to the behavior near the singularity, that is, for cosmic times
$t\simeq t_s$. By substituting the Hubble rate of Eq. (\ref{hublawsing}) in
Eq. (\ref{r3}), results to the following linear second order differential
equation,
\begin{equation}
\label{ptdiffeqn}
2\frac{\mathrm{d}^2P(t)}{\mathrm {d}t^2}-2\left( c_0+f_0(t-t_s)^{\alpha
}\right)\frac{\mathrm{d}P(
t)}{\mathrm{d}t}-4 f_0 (t-t_s)^{-1+\alpha } \alpha P(t)=0\, .
\end{equation}
By making use of the variable $x=t-t_s$, the differential equation
(\ref{ptdiffeqn}) can be written as follows,
\begin{equation}
\label{dgfere}
2\frac{\mathrm{d}^2P(x)}{\mathrm {d}x^2}-2\left( c_0+f_0x^{\alpha
}\right)\frac{\mathrm{d}P(x)}{\mathrm{d}x}-2 f_0 x^{-1+\alpha } \alpha
P(x)=0\, .
\end{equation}
Solving the differential equation (\ref{dgfere}) analytically for a general
$\alpha$, is a rather formidable task, so we shall be interested in the
behavior of the solution for $x\rightarrow 0$, which corresponds to
$t\rightarrow t_s$. So since we assumed $c_0\gg f_0$ and for $x\rightarrow
0$, the differential equation (\ref{dgfere}) becomes,
\begin{equation}
\label{dgferemod}
2\frac{\mathrm{d}^2P(x)}{\mathrm
{d}x^2}-2c_0\frac{\mathrm{d}P(x)}{\mathrm{d}x}=0\, ,
\end{equation}
the solution of which is,
\begin{equation}
\label{genrealsol}
P(x)= \frac{ \e^{c_0 x}  c_1}{1+c_0}+c_2\, ,
\end{equation}
where $c_1$, $c_2$ are arbitrary constants. Since $x\rightarrow 0$, it
stands to reason to expand $P(x)$ for $x\rightarrow 0$, so we obtain,
\begin{equation}\label{pxfap}
P(x)\simeq \frac{c_1}{1+c_0}(1+c_0 x)
\end{equation}

Then, by substituting $P(x)$ in Eq. (\ref{r2}), we get the function $Q(x)$,
which for small $x$ is approximately equal to,
\begin{align}
\label{qtanalyticform}
Q(x) =-6c_0^2c_2-\frac{12 c_0^2c_1}{1+c_0}\left(1+c_0
x+\frac{c_0^2}{2}x^2\right)\, .
\end{align}
Note that we made use of the fact that $c_0\gg 1$. Then, by substituting the
resulting forms of $P'(x)$ and $Q'(x)$ in Eq. (\ref{auxiliaryeqns}), and
solving it with respect to $x$, we get the function $x(R)$. By doing so, we
finally obtain,
acquire
\begin{equation}
\label{finalxr}
x\simeq \frac{A\,R+B}{C} \, ,
\end{equation}
where the detailed form of the parameters $A$, $B$ and $C$ are given in
Appendix. Then by combining Eqs. (\ref{pxfap}), (\ref{qtanalyticform}) and
(\ref{finalxr}) and upon substitution in Eq. (\ref{r1}), we obtain the final
form of the $F(R)$ gravity near the Type IV singularity $t=t_s$, which is,
\begin{equation}
\label{finalfrgravity}
F(R)\simeq R+a_2 R^2+a_0\, ,
\end{equation}
with the detail form of the parameters $a_0$ and $a_2$ appearing in the
Appendix. Note additionally that we have set $c_1=\frac{1+c_0}{4}$, so that
the coefficient of $R$ in Eq. (\ref{finalfrgravity}) becomes equal to one,
and therefore we can have Einstein gravity plus higher curvature terms.

\section{Singular Inflation Analysis and Instabilities for the Inflation Toy
Model}

The inflationary evolution described by the Hubble rate of Eq.
(\ref{hublawsing}) provides the same physical picture that standard
inflation gives. Specifically, during the inflationary era, the cosmological
evolution is a nearly de Sitter evolution, so an exponential expansion
occurs, and the scale factor is of the form $a(t)\sim e^{c_0t}$. More
importantly, the comoving Hubble radius $R_{H}=\frac{1}{a(t)H(t)}$ shrinks
during inflation, and expands after inflation. Moreover, the Type IV
singularity has no particular effect on the comoving quantities, like the
comoving Hubble radius. This remark is very important and this is due to the
presence of the parameter $c_0$. If this was not present, then the standard
inflationary picture would not hold true anymore, since a singularity would
appear in the comoving Hubble radius. In a future work related to bouncing
$F(R)$ cosmology, we shall address this issue in detail.

Coming back to the inflationary evolution (\ref{hublawsing}), the dynamics
of the $F(R)$ gravity cosmological evolution is determined by the Hubble
flow parameters (also known as slow-roll parameters) given below
\cite{encyclopedia,noh},
\begin{equation}\label{hubbleflowpar}
\epsilon_1=-\frac{\dot{H}}{H^2},\,\,\,\epsilon_3=\frac{\sigma
'\dot{R}}{2H\sigma },
\,\,\,\epsilon_4=\frac{\sigma''(\dot{R})^2+\sigma ' \ddot{R}}{H\sigma '
\dot{R}}\, ,
\end{equation}
where $\sigma =\frac{\mathrm{d}F}{\mathrm{d}R}$ and the prime in the above
equation denotes differentiation with respect to the Ricci scalar $R$. The
Hubble flow parameters of Eq. (\ref{hubbleflowpar}) are indicators of the
dynamical evolution of the cosmological system and practically they control
the dynamics. Particularly, if $\epsilon_i\ll 1$, inflation actually occurs,
and it stops when these become of order $\sim 1$. When these become of order
one, the solution of the cosmological dynamical system that described the
evolution up to that point, ceases to be the final attractor of the
dynamical system, so the system exits from the inflationary era.

Let us calculate the Hubble flow parameters of Eq. (\ref{hubbleflowpar}),
for the Hubble rate (\ref{hublawsing}) and for the $F(R)$ gravity of Eq.
(\ref{finalfrgravity}), emphasizing to their form near the Type IV
singularity at $t=t_s$. The parameter $\epsilon_1$ for the Hubble rate
(\ref{hublawsing}) reads,
\begin{equation}\label{hubserd}
\epsilon_1=-\frac{f_0 (t-t_s)^{-1+\alpha } \alpha }{\left(c_0+f_0
(t-t_s)^{\alpha }\right)^2}\, ,
\end{equation}
so near the Type IV singularity, this reads,
\begin{equation}\label{nearthesing}
\epsilon_1\simeq -\frac{f_0 (t-t_s)^{-1+\alpha } \alpha }{c_0^2}\, ,
\end{equation}
and since $\alpha>1$ and $\frac{f_0}{c_0}\ll 1$, we conclude that near the
Type IV singularity the parameter $\epsilon_1$ satisfies $\epsilon_1\ll 1$,
while at the Type IV singularity it becomes equal to zero. The full analytic
form of the parameter $\epsilon_3$ for the Hubble rate (\ref{hublawsing})
and for the $F(R)$ gravity of Eq. (\ref{finalfrgravity}), is equal to,
\begin{equation}\label{dgfffdfire}
\epsilon_3=\frac{6 a_2 f_0 (t-t_s)^{-2+\alpha } \alpha  \left(-1+4 c_0
(t-t_s)+4 f_0 (t-t_s)^{1+\alpha }+\alpha \right)}{\left(c_0+f_0
(t-t_s)^{\alpha }\right) \left(a_1+12 a_2 \left(2 \left(c_0+f_0
(t-t_s)^{\alpha }\right)^2+f_0 (t-t_s)^{-1+\alpha } \alpha \right)\right)}\,
,
\end{equation}
which can be approximated near the Type IV singularity as follows,
\begin{equation}\label{epsilon3forrsquare}
\epsilon_3\simeq \frac{6 a_2 c_0^{-\alpha } f_0 (-1+\alpha ) \alpha }{1+24
a_2 c_0^2}(t-t_s)^{-2+\alpha }\, .
\end{equation}
Finally, the full analytic form of the parameter $\epsilon_4$ appears in the
Appendix, while below we quote its approximate form near the Type IV
singularity,
\begin{equation}\label{epsilon4forrsquare}
\epsilon_4\simeq \frac{2-3 \alpha +\alpha ^2}{c_0 (t-t_s) (-1+\alpha )}\, .
\end{equation}
Let us examine the behavior of the parameters $\epsilon_3$ and $\epsilon_4$
as functions of the cosmic time, starting with $\epsilon_3$. As it is
obvious from Eq. (\ref{epsilon3forrsquare}), the parameter $\epsilon_3$ for
$t<t_s$, satisfies $\epsilon_3\ll 1$, since we assumed that
$\frac{f_0}{c_0}\ll 1$, and also since the cosmic time takes very small
values. However at $t=t_s$, the parameter $\epsilon_3$ blows up, since the
term $\sim (t-t_s)^{-2+\alpha }$, is singular at $t=t_s$, for $1<\alpha<2$.
However, if $\alpha>2$, the parameter $\epsilon_3$ is free of singularity,
but what it interests us the most is the case that a singularity occurs,
that is when $1<\alpha<2$, for which we have a Type IV singularity as we
discussed earlier. The same applies for the parameter $\epsilon_4$ and since
$c_0\gg 1$, for $t<t_s$, we have $\epsilon_4\ll 1$, while at the Type IV
singularity, that is at $t=t_s$, the parameter $\epsilon_4$ diverges, due to
the presence of the term $\sim (t-t_s)$. In this case however, a singularity
occurs regardless what the value of $\alpha$ is.

We are therefore confronted with the following physical picture: The Hubble
flow parameters that control the dynamics of the inflationary solution at
hand, for times near but before the singularity $t<t_s$ take very small
values $\epsilon_i\ll 1$, but at the Type IV singularity, two of these
diverge, for $1<\alpha<2$. This infinite singularity for the parameters
$\epsilon_3$ and $\epsilon_4$ clearly indicates a dynamical instability of
the cosmological system. In addition, we could say that the dynamical
evolution is interrupted violently, so in practise this could mean that the
inflationary solution that described the dynamical cosmological evolution up
to that point, ceases to be the final attractor of the dynamical
cosmological system. Therefore, we could loosely state that this singularity
could be an indicator that inflation ends, and therefore it indicates the
graceful exit from inflation. In conclusion, we demonstrated that the Hubble
flow parameters develop an infinite instability at the Type IV singularity,
and this instability is an indication that graceful exit from inflation is
triggered. Of course, this is just an indication of graceful exit, and in
principle someone could investigate with other methods, the existence of
unstable de Sitter vacua near the Type IV singularity. We perform this study
in the next section for the inflationary toy model.

\subsection{Graceful Exit via de Sitter Vacuum Instability}

Even for the approximate $F(R)$ gravity solution, we can investigate the
stability of the corresponding de Sitter vacua near the Type IV singularity.
This will validate our claim that the graceful exit indeed occurs near the
Type IV singularity.

So we investigate which de Sitter vacua does the $F(R)$ gravity of Eq.
(\ref{finalfrgravity}) has, and whether these are stable or unstable. Let us
recall the fact that if a de Sitter vacuum is unstable towards linear
perturbations, then the solution corresponding to the de Sitter vacuum is
unstable, and therefore it cannot be the final attractor of the theory
\cite{sergeitraceanomaly}. Since the parameters $a_0$ and $a_2$ appearing in
Eq. (\ref{finalfrgravity}), contain the parameter $c_2$, which is a free
parameter in the theory, we can adjust the value of $c_2$, so that the
$F(R)$ gravity has $H(t)=c_0$ as de Sitter solution, so practically, the
dominating part of the Hubble rate (\ref{hublawsing}) is the de Sitter
solution. Since $c_0\gg 1$, the parameter $a_0$ is approximately equal to
$a_0\simeq -6c_0c_2$, so we choose $c_2=2$. Then, the $F(R)$ gravity becomes
equal to,
\begin{equation}\label{frdesitter}
F(R)\simeq -12c_0^2+R+a_2R^2\, .
\end{equation}
In order to investigate the existence of de Sitter vacua for the $F(R)$
gravity of Eq. (\ref{frdesitter}), we shall seek for solutions of the form
$H(t)=H_{dS}$ in the first FRW equation,
\begin{equation}\label{frwdesitter}
6H^2=RF'(R)-F(R)-6H^2F'(R)-6H\frac{\mathrm{d}F'(R)}{\mathrm{d}t}\, ,
\end{equation}
and by substituting in Eq. (\ref{frwdesitter}), the solution $H=H_{dS}$ we
obtain the following algebraic equation,
\begin{equation}\label{alghfeqn}
12 c_0^2-18 H_{dS}^2-144 a_2 H_{dS}^4+6 H_{dS}^2 \left(1+24 a_2
H_{dS}^2\right)=0\, ,
\end{equation}
which can be solved to yield $H_{dS}=c_0$. Therefore, the dominant part of
the Hubble rate (\ref{hublawsing}) is the de Sitter vacuum of the $F(R)$
gravity (\ref{frdesitter}), so what follows is to check the stability of
this vacuum towards linear perturbations of the following form,
\begin{equation}\label{perturbation}
H(t)=H_{dS}+\Delta H(t)\, ,
\end{equation}
assuming that the linear perturbation function $\Delta H (t)$, satisfies
$\mid\Delta H(t)\mid\ll 1$. Substituting Eq. (\ref{perturbation}) in the
first FRW equation (\ref{frwdesitter}), and keeping derivatives up to second
order of $\Delta H(t)$, and also linear terms to $\Delta \dot{H}(t)$, we
obtain the following second order differential equation,
\begin{align}\label{ddotdiffeqna}
& 12 c_0^2-12 H_{dS}^2-24 H_{dS} \Delta H (t)+\frac{27 H_{dS}^2}{4
c_0^2}\Delta\dot{ H} (t)+\frac{9 H_{dS} }{4 c_0^2}\Delta\ddot{ H} (t)=0\, ,
\end{align}
where the ``dot'' denotes differentiation with respect to $t$. Solving the
differential equation (\ref{ddotdiffeqna}), with $H_{dS}=c_0$ we obtain the
following solution,
\begin{equation}\label{solutionbigfr3}
\Delta H(t)=c_Ae^{\frac{1}{6} \left(-9 c_0-\sqrt{465} c_0\right)
t}+c_Be^{\frac{1}{6} \left(-9 c_0+\sqrt{465} c_0\right) t}\, ,
\end{equation}
where $c_A$ and $c_B$ are arbitrary integration constants. Obviously the
second term in the above equation, namely $\sim e^{\frac{1}{6} \left(-9
c_0+\sqrt{465} c_0\right) t}$, grows exponentially with time, since the
exponent is positive, and therefore it dominates the evolution of the linear
perturbation $\Delta H(t)$. Therefore, since the evolution of linear
perturbations $\Delta H(t)$ grows exponentially, this means that the de
Sitter vacuum $H_{dS}=c_0$ is unstable towards linear perturbations, and
therefore this results to curvature perturbations which will induce the
graceful exit from inflation in the present $F(R)$ model. Recall that the
$F(R)$ gravity of Eq. (\ref{frdesitter}) has this form only near the Type IV
singularity, so the analysis holds true for cosmic times near the
singularity, and therefore graceful exit occurs at exactly the Type IV
singularity, since $H(t)=c_0$ at $t=t_s$. This analysis of stable-unstable
de Sitter vacua validates our claim that the infinite dynamical
instabilities in the Hubble flow parameters indicate that graceful exit from
inflation will occur.

\subsection{The Role of the $F(R)$ Gravity}

From the analysis we performed in the previous section, it is obvious that
the functional form of the $F(R)$ gravity plays a crucial role in the
determination of the final form of the Hubble flow parameters, as it can
also be seen from their analytic form given in Eq. (\ref{hubbleflowpar}). It
is worth examining the general case of an $F(R)$ gravity, to see how the
form of the $F(R)$ gravity affects the singularity structure of the Hubble
flow parameters. This is the subject of this section, where we assume a
general form for the $F(R)$ gravity. We shall adopt the following notation
$F_R=F'(R)$, $F_{RR}=F''(R)$, and $F_{RRR}=F'''(R)$.  As it is expected, the
functional form of the $F(R)$ gravity plays an important role on the
singularity structure of the Hubble flow parameters (\ref{hubbleflowpar}).
Actually, under certain conditions, the singularities that the Hubble
flow parameters develop at the Type IV singularity can be avoided.

For a general $F(R)$ gravity, the first Hubble flow parameter $\epsilon_1$
remains unaffected, but the parameters $\epsilon_3$ behaves as follows,
\begin{equation}\label{epsilon3general}
\epsilon_3=\frac{3 f_0 F_{RR} (t-t_s)^{-2+\alpha } \alpha  \left(-1+4 c_0
(t-t_s)+4 f_0 (t-t_s)^{1+\alpha }+\alpha \right)}{F_R \left(c_0+f_0
(t-t_s)^{\alpha }\right)}\, ,
\end{equation}
which for $t\rightarrow t_s$, is simplified as follows,
 \begin{equation}\label{dgdgapprxoima}
\epsilon_3\simeq \frac{3 f_0 F_{RR} (t-t_s)^{-2+\alpha } (-1+\alpha ) \alpha
}{c_0 F_{R}}\, .
\end{equation}
As it can be seen, for the case of a Type IV singularity, and when $1<\alpha
<2$, the parameter $\epsilon_3$ diverges, due to the presence of the term
$\sim (t-t_s)^{-2+\alpha }$. However, if the $F(R)$ gravity behaves as,
\begin{equation}\label{behaviornonsingulari}
\frac{F_{RR}}{F_{R}}\sim (t-t_s)^{\beta }\, ,
\end{equation}
with $\beta \geq (\alpha -2)$, then the Hubble flow parameter $\epsilon_3$
behaves as $\epsilon_3 \sim (t-t_s)^{\beta +\alpha -2}$, so the parameter
$\epsilon_3$ is free of singularities. With regards to the Hubble flow
parameter $\epsilon_4$, for a general $F(R)$ gravity, and for the Hubble
parameter (\ref{hubbleflowpar}), we calculated it's general form and it
appears in the Appendix. Near the Type IV singularity, the parameter
$\epsilon_4$ can be approximated as follows,
\begin{equation}\label{fhfhapproximatee4}
\epsilon_4 \simeq \frac{2-6 \alpha +\alpha ^2}{c_0 (t-t_s) (-1+\alpha
)}+\frac{6 f_0 F_{RRR} (t-t_s)^{-2+\alpha } \left(\alpha -2 \alpha ^2+\alpha
^3\right)}{c_0 F_{RR} (-1+\alpha )}\, .
\end{equation}
As in the case of the parameter $\epsilon_3$, the parameter $\epsilon_4$
diverges for $1<\alpha<2$, unless the $F(R)$ gravity behaves as follows,
\begin{equation}\label{e3e4nondivergence}
\frac{F_{RRR}}{F_{RR}}\sim (t-t_s)^{\beta}\, ,
\end{equation}
with $\beta \geq \alpha-2$, as in the previous case. It is remarkable that
both the Hubble flow parameters $\epsilon_3$ and $\epsilon_4$ are free of
singularities if the functional form of the $F(R)$ gravity satisfies similar
functional relations, namely Eq. (\ref{behaviornonsingulari}) and
(\ref{e3e4nondivergence}), and in both cases for $\beta \geq \alpha -2$.

In conclusion, the functional form of the $F(R)$ gravity plays an important
role on the singularity structure of the Hubble flow parameters, and
therefore on the potential dynamical instability of the cosmological system
under study. Particularly, the singularities occur if $\alpha$ takes values
in the interval $(1,2)$, however, if the $F(R)$ gravity satisfies,
\begin{equation}\label{dsijfgdbf}
\frac{F_{RRR}}{F_{RR}}\sim \frac{F_{RR}}{F_{R}}\sim (t-t_s)^{\beta }\, ,
\end{equation}
with $\beta \geq \alpha -2$, then the Hubble flow parameters are free of singularities.

\subsection{The Hubble Slow-Roll Parameters for a Type IV Singular
Evolution}

In the previous sections we used the Hubble flow parameters of Eq.
(\ref{hubbleflowpar}) to show that an instability occurs in the case the
cosmological evolution of a Jordan frame $F(R)$ gravity contains a Type IV
singularity. This instability was actually an infinite instability, since
some of the Hubble flow parameters contain singularities that occur at the
Type IV singularity. As we discussed earlier, the Hubble flow parameters
determine the dynamics of inflation, and their functional form depends
strongly on the functional form of the $F(R)$ gravity. However, in the
literature there are also other parameters that also determine the dynamics
of inflation, and in this section we shall discuss the Hubble slow-roll
parameters $\epsilon_H$ and $\eta_H$ \cite{barrowslowroll}. The Hubble
slow-roll parameters are superior to the standard slow-roll parameters
\cite{barrowslowroll}, but when these are calculated for a canonical scalar
field. In the present case, if someone uses the Hubble slow-roll parameters,
it is less transparent how the form of the $F(R)$ gravity modifies the
dynamics of inflation, but we shall calculate these for completeness. We
need to stress however, that the Hubble flow parameters offer a much more
reliable description to the dynamics of inflation, at least in the case that
a Jordan frame $F(R)$ gravity is studied.

The Hubble slow-roll parameters $\epsilon_H$ and $\eta_H$ are equal to
\cite{barrowslowroll},
\begin{equation}\label{hubbleslowrooll}
\epsilon_H=-\frac{\dot{H}}{H^2},\,\,\, \eta_H=-\frac{\ddot{H}}{2H\dot{H}}\,
,
\end{equation}
so by substituting the Hubble rate of Eq. (\ref{hublawsing}) in Eq.
(\ref{hubbleslowrooll}), the parameter $\epsilon_H$ becomes,
\begin{equation}\label{epsilonforsingularevolution}
\epsilon_H=-\frac{f_0 (t-t_s)^{-1+\alpha } \alpha }{\left(c_0+f_0
(t-t_s)^{\alpha }\right)^2}\, ,
\end{equation}
while the parameter $\eta_H$ is equal to,
\begin{equation}\label{etasingualarhubla}
\eta_H=-\frac{-1+\alpha }{2 \left(c_0+f_0 (t-t_s)^{\alpha }\right)
(t-t_s)}\, .
\end{equation}
Therefore, near the Type IV singularity, the Hubble slow-roll parameters
$\epsilon_H$ and $\eta_H$ of Eqs. (\ref{epsilonforsingularevolution}) and
(\ref{etasingualarhubla}) become,
\begin{equation}\label{nearthesingularity}
\epsilon \simeq -\frac{f_0 (t-t_s)^{-1+\alpha } \alpha }{c_0^2}, \,\,\,
\eta_H \simeq -\frac{-1+\alpha }{2 c_0 (t-t_s)}
\end{equation}
from which we can see that in the case of a Type IV singularity, which
occurs when $\alpha >1$, the first Hubble slow-roll parameter $\epsilon_H$
is regular at the singularity, while the second slow-roll parameter $\eta_H$
blows up at the Type IV singularity. Therefore, even the Hubble slow-roll
parameters have similar behavior for the singular Hubble rate
(\ref{hublawsing}), with some of them exhibiting an infinite instability.
Since the Hubble slow-roll parameters also govern the inflationary dynamics,
the infinite instability at the singularity clearly indicates that the
solution which described the dynamical evolution up to that point is clearly
not the final attractor of the theory, since the dynamical system is
abruptly interrupted. Therefore, this could be an indicator that graceful
exit from inflation occurs.

Before we end this section, we need to note that the infinite instability
induced by the Type IV singularity could be viewed as another mechanism that
could potentially indicate the graceful exit from inflation. Of course, the
actual mechanism for the graceful exit should be found by searching for any
possible tachyonic instabilities \cite{encyclopedia}, or unstable de Sitter
vacua \cite{sergeitraceanomaly}, or even for possible quantum effects
\cite{Nojiri:2005sx}, all these related somehow with the Type IV
singularity. So the infinite instabilities of the Hubble flow parameters are
actually indicators that graceful exit occurs. In this paper, we used the
second mechanism related to unstable de Sitter vacua, to show that indeed
graceful exit occurs near the singularity. As we demonstrated, an unstable
de Sitter solution exists near the Type IV singularity, so this validated
our claims to some extend. However, what is lacking is a detailed study of
the $F(R)$ gravity dynamical system near the singularity, which we hope to
address in a future work.

\section{Singular Inflation and Comparison with Ordinary Starobinsky
Inflation}

\section{Non-Singular Starobinsky Inflation}

Although the functional form of the $F(R)$ gravity (\ref{finalfrgravity})
resembles very much the ordinary Starobinsky $R^2$ inflation model \cite{starobinsky}, it
is not the Starobinsky model, as it can be seen from the parameters $a_0$
and $a_2$ appearing in the Appendix. It is of great importance to
investigate what new qualitative features does the singularity during
inflation brings along. In order to do so, we shall study the $R^2$ inflation
model, with a singularity being included and compare our results with the
ordinary $R^2$ inflation model. This is necessary in order to understand the new
qualitative features of the singular inflation. To start with, let us
present the ordinary $R^2$ inflation model, which we modify later on in order to
include a Type IV singularity. In the following, when we mention ``ordinary
$R^2$ inflation model'', we mean the non-singular version of the Starobinsky $R^2$ inflation model. For the
$R^2$ inflation model, the $F(R)$ gravity is,
\begin{equation}\label{starobfr}
F(R)=R+\frac{1}{6M^2}R^2\, ,
\end{equation}
with $M\gg 1$. The FRW equation corresponding to the $F(R)$ gravity
(\ref{starobfr}) is given below,
\begin{equation}\label{takestwo}
\ddot{H}-\frac{\dot{H}^2}{2H}+\frac{M^2}{2}H=-3H\dot{H}\, ,
\end{equation}
and since during inflation, the terms $\ddot{H}$ and $\dot{H}$ can be
neglected, the resulting Hubble rate that describes the $R^2$ inflation model of Eq. (\ref{starobfr})
is,
\begin{equation}\label{hubstar}
H(t)\simeq H_i-\frac{M^2}{6}\left ( t-t_i\right )\, .
\end{equation}
with $t_i$ the time instance that inflation starts and also $H_i$ the value
of the Hubble rate at $t_i$. Let us calculate the Hubble flow parameters for
the ordinary $R^2$ inflation model of Eq. (\ref{starobfr}), which we will need later in order to compare
with the singular version. By substituting Eqs. (\ref{hubstar}) and
(\ref{starobfr}) in Eq. (\ref{hubbleflowpar}), the Hubble flow parameters
for the $R^2$ inflation model of Eq. (\ref{starobfr}) model become,
\begin{align}\label{starobhubflow}
& \epsilon_1=\frac{M^2}{6 \left(H_i-\frac{1}{6} M^2 (t-t_i)\right)^2}, \\
\notag &
\epsilon_3= -\frac{2}{3 \left(1+\frac{2 \left(-\frac{M^2}{6}+2
\left(H_i+\frac{1}{6} M^2 (-t+t_i)\right)^2\right)}{M^2}\right)}, \\ \notag
& \epsilon_4=-\frac{M^2}{6 \left(H_i-\frac{1}{6} M^2 (t-t_i)\right)^2}\, ,
\end{align}
from which we can see that no singularity occurs, as it was expected.

In the $R^2$ inflation model of Eq. (\ref{starobfr}), inflation ends when the Hubble flow parameter
$\epsilon_1$ becomes $\epsilon_1\sim 1$, so in principle the graceful exit
comes as a result of the breaking of the slow-roll expansion
\cite{barrowslowroll}. Indeed, the slow-roll expansion is a perturbative
expansion in terms of the Hubble flow parameters, and it is valid until the
time instance that the perturbative approximation breaks down. In the ordinary
$R^2$ inflation model, the perturbative expansion breaks when the first
slow-roll parameter $\epsilon_1$ becomes $\epsilon_1\sim 1$, which occurs at
the time instance $t_f$. The Hubble rate at $t_f$ is
$H_f=Hi-\frac{M^2}{6}(t_f-t_i)$, which since $\epsilon_1\simeq 1$ at
$t=t_f$, this means that, $H_f\simeq \frac{M}{\sqrt{6}}$. It is very
important to see how the rest Hubble flow parameters behave at $t=t_f$, and
by substituting the value of $H_f$ in $\epsilon_3$ and $\epsilon_4$ we get,
\begin{equation}\label{nonsinginfla}
\epsilon_3\simeq -\frac{1}{2},\,\, \,\,\epsilon_4 \simeq -1\, .
\end{equation}
Also, the time instance $t_f$ that inflation ends is equal to,
\begin{equation}\label{tfending}
t_f\simeq t_i+\frac{6H_i}{M^2}\, ,
\end{equation}
which easily follows from the condition $\epsilon_1(t_f)\simeq 1$. In
principle, as it is known from the literature \cite{barrowslowroll},
inflation ends when the perturbative slow-roll expansion breaks down. If
this occurs, then the solution that described the dynamical evolution of the
inflationary model up to that point, ceases to be the final attractor of the
model. It is conceivable that the first Hubble flow parameter $\epsilon_1$
corresponds to the first order term in the slow-roll expansion, and as it
was argued in \cite{barrowslowroll}, it is possible that graceful exit might
occur before this term becomes of order one, since the higher order Hubble
flow parameters might become non-perturbative before $\epsilon_1$ becomes of
order one. However, in the ordinary $R^2$ inflation model this does not occur,
since, as it can be seen from Eq. (\ref{nonsinginfla}), the rest of the
Hubble flow parameters become of order one simultaneously with $\epsilon_1$.
As we demonstrate in the next section, this is the difference between the
singular $R^2$ inflation and the ordinary $R^2$ inflation.

Before we close this section, it is worth calculating the Hubble slow-roll
indices (\ref{hubbleslowrooll}) for the ordinary $R^2$ inflation model, and also
express these in term of the $e-$folds number $N$, which is defined as
follows,
\begin{equation}\label{efoldings}
N=\int_{t_i}^{t}H(t)\mathrm{d}t\, .
\end{equation}
The spectral index of primordial curvature perturbations $n_s$ and the
scalar-to-tensor ratio in terms of the Hubble slow-roll parameters $\eta_H$
and $\epsilon_H$ are equal to \cite{barrowslowroll},
\begin{equation}\label{egefda}
n_s\simeq 1-4\epsilon_H +2\eta_H,\, \, \,r=48 \epsilon_H^2 \, ,
\end{equation}
which holds true only in the case the slow-roll expansion is valid. This is
a very important observation, since if one of the Hubble slow-roll
parameters is large enough so that the slow-roll expansion breaks down, then
the observational indices are not given by Eq. (\ref{egefda}).

Assuming that the Hubble slow-roll parameters are such, so that the
slow-roll approximation holds true, let us calculate the Hubble slow-roll
parameters and inflationary indices for the Hubble rate (\ref{hubstar}). The
Hubble slow-roll indices read,
\begin{equation}\label{hubslowordstarob}
\epsilon_H=\frac{M^2}{6 \left(H_i-\frac{1}{6} M^2 (t-t_i)\right)^2},\,\,\,
\eta_H=0\, .
\end{equation}
We can express the Hubble slow-parameter $\epsilon_H$ in term of $N$, and by
combining Eqs. (\ref{efoldings}) and (\ref{hubstar}), we obtain,
\begin{equation}\label{ggdfvgvhsdgd}
t-t_i=\frac{2 \left(3 H_i+\sqrt{3} \sqrt{3 H_i^2-M^2 N}\right)}{M^2}\, ,
\end{equation}
so upon substitution in Eq. (\ref{hubslowordstarob}) we get,
\begin{equation}\label{epskiolonsokio}
\epsilon_H=\frac{M^2}{6 H_i^2-2 M^2 N}\, .
\end{equation}
Consequently, the spectral index $n_s$ and the scalar-to-tensor ratio $r$,
read,
\begin{equation}\label{gdgfdfdf}
n_s=1-\frac{4M^2}{6 H_i^2-2 M^2 N},\,\,\, r=48\left(\frac{M^2}{6 H_i^2-2 M^2
N}\right)^2\, .
\end{equation}
The recent observations of the Planck collaboration \cite{planck} have
verified that the $R^2$ inflation model is in concordance with observations, so
if we suitably choose $M$ and $H_i$, concordance may be achieved. Of course
our approach is based on a Jordan frame calculation, but the resulting
picture with regards to the observational indices is the same in both Jordan
and Einstein frame \cite{kaizer}. To be more concrete, let us see for which values of $H_i$, $M$ and $N$ we can achieve concordance with observations. Assume for example that the number of $e$-folds is $N=60$, so for $M\sim 10^{13}$$\mathrm{sec}^{-1}$, and $H_i\sim 6.29348 \times 10^{13}$$\mathrm{sec}^{-1}$, we obtain that the spectral index of primordial perturbations $n_s$ and the scalar-to-tensor ratio $r$ become approximately,
\begin{equation}\label{scealertensorbfer}
n_s\simeq 0.966,\,\,\, r\simeq 0.003468\, .
\end{equation}
The latest Planck data (2015) \cite{planck} indicate that $n_s$ and $r$ are approximately equal to,
\begin{equation}\label{recentplancdata}
n_s=0.9655\pm 0.0062\, , \quad r<0.11\, ,
\end{equation}
so the values given in Eq. (\ref{scealertensorbfer}) are in concordance with the current observational data.

\subsection{Singular Inflation}

The ordinary $R^2$ inflation can also contain a Type IV singularity
that we assume to occur at $t=t_s$. The Hubble rate that will describe the
singular inflation evolution is the following,
\begin{equation}\label{singstarobhub}
H(t)\simeq H_i-\frac{M^2}{6}\left ( t-t_i\right )+f_0\left (t-t_s
\right)^{\alpha} \, ,
\end{equation}
and we shall assume that $\alpha >1 $, so that a Type IV singularity occurs.
In addition, we assume that $H_i\gg f_0$, $M\gg f_0$ and also that $f_0\ll
1$, and consequently the singularity term is significantly smaller in
comparison to the first two terms in Eq. (\ref{singstarobhub}). Hence, at
the Hubble rate level, the singularity term remains small during inflation
and therefore it can be unnoticed. Therefore, near $t\simeq t_s$, the $F(R)$
gravity that can generate the evolution (\ref{singstarobhub}) is the one
appearing in Eq. (\ref{starobfr}). As we demonstrated previously, the
effects of the singularity will not appear at the level of observable
quantities, but the singularity will strongly affect the dynamics of the
system. Now we investigate in detail if this holds true in this case too. In
order to see this, we calculate the Hubble flow parameters for the Hubble
rate (\ref{singstarobhub}), so these read,
\begin{align}\label{singstarohubflow}
& \epsilon_1=-\frac{-\frac{M^2}{6}+f_0 (t-t_s)^{-1+\alpha } \alpha
}{\left(H_i-\frac{1}{6} M^2 (t-t_i)+f_0 (t-t_s)^{\alpha }\right)^2},  \\
\notag &
\epsilon_3= \frac{f_0 (t-t_s)^{-2+\alpha } (-1+\alpha ) \alpha +4
\left(H_i-\frac{1}{6} M^2 (t-t_i)+f_0 (t-t_s)^{\alpha }\right)
\left(-\frac{M^2}{6}+f_0 (t-t_s)^{-1+\alpha } \alpha \right)}{M^2
\left(1+\frac{2 \left(-\frac{M^2}{6}+2 \left(H_i+\frac{1}{6} M^2
(-t+t_i)\right)^2\right)}{M^2}\right) \left(H_i-\frac{1}{6} M^2 (t-t_i)+f_0
(t-t_s)^{\alpha }\right)}\\ \notag &
\epsilon_4=\frac{4 f_0 H(t) (t-t_s)^{-2+\alpha } (-1+\alpha ) \alpha +f_0
(t-t_s)^{-3+\alpha } (-2+\alpha ) (-1+\alpha ) \alpha +4
\left(-\frac{M^2}{6}+f_0 (t-t_s)^{-1+\alpha } \alpha \right)^2}{H(t)
\left(f_0 (t-t_s)^{-2+\alpha } (-1+\alpha ) \alpha +4 H(t)
\left(-\frac{M^2}{6}+f_0 (t-t_s)^{-1+\alpha } \alpha \right)\right)}\, .
\end{align}
which for cosmological times near the singularity at $t=t_s$, these can be
simplified as follows,
\begin{align}\label{singstarohubflow1}
& \epsilon_1=\frac{M^2}{6 \left(H_i-\frac{1}{6} M^2 (t-t_i)\right)^2},  \\
\notag &
\epsilon_3= \frac{f_0 (t-t_s)^{-2+\alpha } (-1+\alpha ) \alpha +4
\left(H_i-\frac{1}{6} M^2 (t-t_i)\right) \left(-\frac{M^2}{6}\right)}{M^2
\left(1+\frac{2 \left(-\frac{M^2}{6}+2 \left(H_i+\frac{1}{6} M^2
(-t+t_i)\right)^2\right)}{M^2}\right) \left(H_i-\frac{1}{6} M^2
(t-t_i)\right)} \\ \notag &
\epsilon_4=\frac{\frac{M^4}{9}+4 f_0 \left(H_i-\frac{1}{6} M^2
(t-t_i)\right) (t-t_s)^{-2+\alpha } (-1+\alpha ) \alpha +f_0
(t-t_s)^{-3+\alpha } (-2+\alpha ) (-1+\alpha ) \alpha
}{\left(H_i-\frac{1}{6} M^2 (t-t_i)\right) \left(-\frac{2}{3} M^2
\left(H_i-\frac{1}{6} M^2 (t-t_i)\right)+f_0 (t-t_s)^{-2+\alpha } (-1+\alpha
) \alpha \right)}\, .
\end{align}
Note that in Eq. (\ref{singstarohubflow}) $H(t)$ is the Hubble rate
(\ref{singstarobhub}). As it can be seen from Eq. (\ref{singstarohubflow1}),
the Hubble flow parameter $\epsilon_1$ for the singular inflation model, is
identical to the one corresponding to the ordinary $R^2$ inflation model, given
in Eq. (\ref{starobhubflow}). For the case of the parameter $\epsilon_3$,
things are different, since the presence of the term $\sim
(t-t_s)^{-2+\alpha }$ renders the parameter $\epsilon_3$ singular, if
$1<\alpha<2$, at exactly the point $t=t_s$. In addition, the parameter
$\epsilon_4$ contains two sources of singularity, due to the existence of
the terms $\sim (t-t_s)^{-3+\alpha }$ and $(t-t_s)^{-2+\alpha }$, the first
of which becomes singular for $1<\alpha<3$. The physical implications of
these singularities are quite interesting. Let us assume that the parameter
$\epsilon_1$ becomes of order one at the time instance $t=t_f$, given in Eq.
(\ref{tfending}). So we have many different cosmological scenarios which we
describe in the following separately.

\subsubsection{Scenario I}

If $t_s<t_f$, and $2<\alpha <3$, the parameter $\epsilon_4$ becomes singular
at $t=t_s$, and the rest Hubble flow parameters are not singular.
Particularly, in this case, $\epsilon_1$ remains the same as in Eq.
(\ref{starobhubflow}), while $\epsilon_3$ becomes simplified and behaves as,
\begin{equation}\label{evasd}
\epsilon_3\simeq -\frac{2}{3 \left(1+\frac{2 \left(-\frac{M^2}{6}+2
\left(H_i+\frac{1}{6} M^2 (-t+t_i)\right)^2\right)}{M^2}\right)}\, ,
\end{equation}
which is identical to the one appearing in Eq. (\ref{starobhubflow}) which
corresponds to the ordinary $R^2$ inflation model. Therefore, only the parameter
$\epsilon_4$ remains singular at $t=t_s$, and takes the following form,
\begin{equation}\label{epsilon4sdforscenar}
\epsilon_4\simeq -\frac{3 \left(\frac{M^4}{9}+f_0 (t-t_s)^{-3+\alpha }
(-2+\alpha ) (-1+\alpha ) \alpha \right)}{2 M^2 \left(H_i-\frac{1}{6} M^2
(t-t_i)\right)^2}\, .
\end{equation}

The Hubble flow parameters control the slow-roll expansion
\cite{barrowslowroll}, so a singularity at a higher order slow-roll
parameter indicates a dynamical instability of the system. Actually, it
indicates that at higher orders, the slow-roll perturbative expansion breaks
down, and therefore this indicates that the solution describing the
dynamical evolution of the cosmological system up to that point, ceases to
be an attractor of the system. This clearly may be viewed as a mechanism for
graceful exit from inflation, at least at a higher order.

In conclusion, we have the following qualitative picture: If the Type IV
singularity occurs earlier than the time instance $t_f$ at which
$\epsilon_1$ becomes $\sim 1$, and also if $2<\alpha <3$, then the higher
order Hubble flow parameter becomes singular at the Type IV singularity, and
therefore the higher order Hubble parameters become singular ($\epsilon_4$).
We interpret this infinite instability as an indication that graceful exit
occurs at the singularity, so in this case, the exit comes earlier from the
time that the parameter $\epsilon_1$ becomes of the order $\sim 1$.

It is worth calculating the spectral index of primordial curvature
perturbations $n_s$ and the scalar-to-tensor ratio $r$ in this case, in
order to see how the occurrence of a singularity at $t_s$ affects the
observational data. In order to calculate the inflationary indices, we shall
calculate the Hubble slow-roll parameters $\epsilon_H$ and $\eta_H$ of Eq.
(\ref{hubbleslowrooll}) for the present case. So at $t=t_s$, these
parameters read,
\begin{align}\label{hubslowatts}
& \epsilon_H(t_s)=\frac{M^2}{6 \left(H_i-\frac{1}{6} M^2
(t_s-t_i)\right)^2},\,\,\, \eta_H(t_s)=0\, .
\end{align}
The $e$-folds parameter $N$ of Eq. (\ref{efoldings}), for the Hubble rate
(\ref{singstarobhub}), integrated in the interval $[t_i,t_s]$ becomes,
\begin{equation}\label{nforsingscenarioi}
N=\int_{t_i}^{t_s}H(t)\mathrm{d}t=H_i(t_s-t_i)-\frac{M^2}{12}\left(
t_s-t_i\right)^2\, ,
\end{equation}
so upon solving with respect to $(t_s-t_i)$, we get,
\begin{equation}\label{tsti}
t_s-t_i=\frac{2 \left(3 H_i+\sqrt{3} \sqrt{3 H_i^2-M^2 N}\right)}{M^2}\, ,
\end{equation}
and upon substitution to Eq. (\ref{hubslowatts}), we obtain,
\begin{align}\label{hubslowatt123s}
& \epsilon_H(t_s)=\frac{M^2}{6 H_i^2-2 M^2 N},\,\,\, \eta_H(t_s)=0\, ,
\end{align}
so the Hubble slow-roll parameters are identical to the ones corresponding
to the $R^2$ inflation model, given in Eq. (\ref{epskiolonsokio}). So the
present evolution scenario is quite appealing, the Hubble slow-roll
parameters are identical to the ones corresponding to the ordinary
$R^2$ inflation model. In addition, since the Hubble slow-roll indices
(\ref{hubslowatt123s}) are small during inflation, the observational indices
are also identical to the ones of the ordinary $R^2$ inflation model, so in this
case too we have,
\begin{equation}\label{gdgfdfdfsceanrioI}
n_s=1-\frac{4M^2}{6 H_i^2-2 M^2 N},\,\,\, r=48\left(\frac{M^2}{6 H_i^2-2 M^2
N}\right)^2\, .
\end{equation}
Obviously, concordance with the observations can be achieved, like in the ordinary $R^2$ inflation model. For example, if we assume that the total number of $e$-folds is $N=55$, and also by choosing $M\sim 10^{13}$$\mathrm{sec}^{-1}$ and $H_i\sim 6.15964 \times 10^{13}$$\mathrm{sec}^{-1}$, the spectral index of primordial curvature perturbations $n_s$ and the scalar-to-tensor ratio become,
\begin{equation}\label{scealertensorbfernewcase12344i}
n_s\simeq 0.966,\,\,\, r\simeq 0.003468\, ,
\end{equation}
as in the ordinary $R^2$ inflation model, so comparing with the observational data (\ref{recentplancdata}), it can be seen than concordance can be achieved. Note that we chose $N=55$, since in the case at hand, inflation ends earlier than in the ordinary $R^2$ inflation model.

The differences of the singular inflation compared to the $R^2$ inflation model is that inflation ends
earlier than the $R^2$ inflation model, and also, inflation ends abruptly, since
the Hubble flow parameter $\epsilon_4$ severely diverges. A last comment is
in order: Note that, since this result we obtained for this scenario, holds
for cosmic times in the vicinity of the singularity, so near $t\sim t_s$,
hence it is valid only near the singularity. In principle, the singularity
can be chosen arbitrarily, but then the $e$-folding number should be
appropriately changed. In order to obtain $N\simeq 50-60$, we assume that
$t_s$ is near the cosmic time $t_f$. The most important feature of this
cosmological scenario is that inflation ends abruptly, compared to the
ordinary $R^2$ inflation model, and in fact it ends before the first Hubble
slow-roll parameter becomes of order $\sim 1$. Recall that the first Hubble
slow-roll parameter corresponds to first order in the slow-roll
approximation, so in the present scenario, inflation ends at a higher order
in the slow-roll expansion. We need to note that in this case, the
singularity will not have any observational implications, since the indices
are the same as in the $R^2$ inflation case, with different $N$, $H_i$ and $M$
of course. The only new feature that this scenario brings along is that
inflation seems to end earlier and more abruptly.

\subsubsection{Scenario II}

Consider now the case $t_s<t_f$, and $1<\alpha<2$, in which case the
parameters $\epsilon_3$, $\epsilon_4$ are given in Eq.
(\ref{singstarohubflow1}), while the second Hubble slow-roll parameter
$\eta_H$ is equal,
\begin{equation}\label{gdhdtf}
\eta_H\simeq -\frac{f_0 (t-t_s)^{-2+\alpha } (-1+\alpha ) \alpha }{2
\left(H_i-\frac{1}{6} M^2 (t-t_i)+f_0 (t-t_s)^{\alpha }\right)
\left(-\frac{M^2}{6}+f_0 (t-t_s)^{-1+\alpha } \alpha \right)}\, .
\end{equation}
In this case, $\epsilon_3$, $\epsilon_4$ and $\eta_H$ diverge at $t=t_s$,
which means that the dynamical evolution develops a strong instability at
$t=t_s$, which can be viewed as an indication that inflation ends abruptly
at that point. Since the second Hubble slow-roll index diverges at $t=t_s$,
someone could claim that the observational indices diverge at $t=t_s$.
However, this is wrong, since both the spectral index of primordial
curvature perturbations and the scalar-to-tensor ratio can be expressed in
terms of the Hubble slow-roll parameters if the slow-roll approximation
holds true, that is $\epsilon_H,\eta_H\ll 1$, and obviously at $t=t_s$, the
slow-roll approximation breaks down, since $\eta_H$ diverges. Hence, for the
present case scenario, what we have is a strong infinite instability at
$t=t_s$ which indicates that the dynamical evolution of the cosmological
system is interrupted abruptly. However, we cannot be sure for any
observational implications of the singularity in this case.

\subsubsection{Scenario III}

In the case that $t_s<t_f$ and $\alpha>3$, the parameters $\epsilon_3$,
$\epsilon_4$ are regular at $t=t_s$, and also the Hubble slow-roll parameter
is equal to zero at $t=t_s$. Therefore, in this case inflation ends when
$\epsilon_1$ becomes of order $\sim 1$, and therefore it is almost similar
to the ordinary $R^2$ inflation model and the singularity is almost unnoticed,
because at $t=t_f$, the second Hubble slow-roll parameter $\eta_H$ is not
exactly zero, as in the ordinary $R^2$ inflation model, but it is equal to,
\begin{equation}\label{fgeddsecondhubinde}
\eta_H\simeq -\frac{f_0 (t_f-t_s)^{-2+\alpha } (-1+\alpha ) \alpha }{2
\left(H_i-\frac{1}{6} M^2 (t_f-t_i)+f_0 (t_f-t_s)^{\alpha }\right)
\left(-\frac{M^2}{6}+f_0 (t_f-t_s)^{-1+\alpha } \alpha \right)}\, ,
\end{equation}
and although $(t_f-t_s)\ll 1$, still the parameter $\eta_H$ is not zero, so
we need to stress the difference in comparison to the ordinary $R^2$ inflation
model. In this case, the inflationary index $n_s$ reads,
\begin{equation}\label{secanrioiiobservindex}
n_s=-\frac{9 f_0 (t_f-t_s)^{-1+\alpha } (-1+\alpha ) \alpha }{\left(-\sqrt{9 H_i^2-3 M^2 N}+3 f_0 (t_f-t_s)^{\alpha }\right) \left(M^2 (-t_f+t_s)+6 f_0 (t_f-t_s)^{\alpha } \alpha \right)}
\, .
\end{equation}
Hence in this case, the Type IV singularity may have observational
implications, since the spectral index of primordial curvature perturbations
is different compared to the ordinary $R^2$ inflation model. Of course the
difference is very small $\sim (t_f-t_s)^{-1+\alpha }$, but still the
difference should be reported. Let us be quantitative at this spot, to see how the Type IV singularity affects the spectral index of primordial curvature perturbations, since the scalar-to-tensor ratio remains completely unaffected. By choosing for example the total number of $e$-folds $N=50$, and also by choosing $M\sim 10^{13}$$\mathrm{sec}^{-1}$ and $H_i\sim 6.15964 \times 10^{13}$$\mathrm{sec}^{-1}$, the spectral index of primordial curvature perturbations $n_s$ and the scalar-to-tensor ratio $r$ become,
\begin{equation}\label{scealertensorbfernewcasei}
n_s\simeq 0.966-9.59 \times 10^{-90} ,\,\,\, r\simeq 0.003468\, ,
\end{equation}
since the contribution of $\eta_H$ in the observational parameter $\eta_H$ is very small and particularly $\eta_H\simeq -4.795 \times 10^{-90}$, for the values of $N,H_i$ and $M$ we chose earlier. Obviously, in this scenario too there is concordance with the observational data of Eq. (\ref{recentplancdata}). Hence, it is highly unlikely that the singularity will have an observable effect, unless $H_i$ and $M$ are not chosen to be so large but are of the same order as $(t_f-t_s)^{-1+\alpha }$. However, the latter case is ruled out by observational data. In Table \ref{newtab} we present the spectral index and scalar-to-tensor ratio corresponding to the scenarios I and III, for $M\sim 10^{13}$$\mathrm{sec}^{-1}$ and $H_i\sim 6.15964 \times 10^{13}$$\mathrm{sec}^{-1}$ and for various values of the $e$-folds number $N$. The scenarios are indistinguishable observationally, but these are different from a dynamical point of view.

\begin{table*}[h]
\small
\caption{\label{newtab}Observational Indices for Scenarios I and III for $M\sim 10^{13}$$\mathrm{sec}^{-1}$ and $H_i\sim 6.15964 \times 10^{13}$$\mathrm{sec}^{-1}$}
\begin{tabular}{@{}crrrrrrrrrrr@{}}
\tableline
\tableline
\tableline
Scenario $\,\,\,$ &$\,\,\,\,\,\,\,\,\,\,\,\,\,\,\,$$e$-folding $N$ & $\,\,\,$Spectral Index $n_s$$\,\,\,$ & Scalar-to-Tensor Ratio $r$
\\\tableline
Scenario I $\,\,\,$ & $50$ &$\,\,\,$$n_s\simeq 0.968664$$\,\,\,$ & $r\simeq 0.00294591$
\\\tableline
Scenario I $\,\,\,$ & $60$ &$\,\,\,$$n_s\simeq 0.962842$$\,\,\,$ & $r\simeq 0.00414226$
\\\tableline
Scenario III $\,\,\,$ & $50$ &$\,\,\,\,\,\,\,\,\,\,\,\,\,\,\,$$\,\,\,\,\,\,\,\,\,\,\,\,\,\,\,$$\,\,\,\,\,\,\,\,\,\,\,\,\,\,\,$  $n_s\simeq 0.966-9.59 \times 10^{-90}$   $\,\,\,$ & $r\simeq 0.003468$
\\\tableline
\tableline
\tableline
 \end{tabular}
\end{table*}

\subsubsection{The Scenarios with $t_s=t_f$}

Now we consider all the cases with $t_s=t_f$, and we start our analysis with
the case $2<\alpha<3$, in which case, inflation ends at $t_s=t_f$ at which
time instance, only the Hubble flow parameter $\epsilon_4$ is divergent at
$t=t_f$. This case is similar to Scenario I, so inflation ends at $t=t_f$,
but more abruptly in comparison to the ordinary $R^2$ inflation model, since in
the present case, at $t=t_s$ the parameter $\epsilon_1$ becomes of the order
$\sim 1$ and at the same time, the parameter $\epsilon_4$ diverges at
$t=t_f=t_s$, so the exit becomes in some sense more pronounced.

The case $t_s=t_f$ and $1<\alpha<2$, is similar to Scenario II, with
$\epsilon_3$, $\epsilon_4$ and $\eta_H$ diverging at $t=t_f=t_s$, which
indicates that inflation actually ends at $t=t_f$ but in a more abrupt way.

The case $t_s=t_f$ and $\alpha>3$ is similar to Scenario III with the
difference that in this case, the second Hubble slow-roll index is zero,
since $t_s=t_f$. Therefore, this case is indistinguishable from the ordinary
$R^2$ inflation model. In Table \ref{TableI} we gathered all our results
corresponding to all the scenarios we discussed above.

\begin{table*}[h]
\small
\caption{\label{TableI}Various Scenarios for the Type IV Singular Inflation
Model}
\begin{tabular}{@{}crrrrrrrrrrr@{}}
\tableline
\tableline
\tableline
Time that Type IV \\ Singularity Occurs$\,\,\,$ &$\,\,\,$ Values of
$\alpha$$\,\,\,$ & Compatible with Observations $\,\,\,$ & Graceful
Exit$\,\,$ & Singular Parameters & Singularity observed?
\\\tableline
$t=t_s<t_f$ & $2<\alpha<3$ &
Yes$\,\,\,\,\,\,\,\,\,\,\,\,\,\,\,\,\,\,\,\,\,\,\,\,\,\,\,\,\,\,\,\,\,\,\,$
& Occurs at $t=t_s$ & $\epsilon_4$
$\,\,\,\,\,\,\,\,\,\,\,\,\,\,\,\,\,\,\,\,\,\,\,\,\,$& No
$\,\,\,\,\,\,\,\,\,\,\,\,\,\,\,\,\,\,\,\,\,$\\
\tableline
$t=t_s<t_f$ & $1<\alpha<2$ & Instability
$\,\,\,\,\,\,\,\,\,\,\,\,\,\,\,\,\,\,\,\,\,\,\,\,\,\,$ & Occurs at $t=t_s$ &
$\epsilon_4$, $\epsilon_3$, $\eta_H$ $\,\,\,\,\,\,\,\,\,\,\,\,\,\,\,\,\,\,$&
No $\,\,\,\,\,\,\,\,\,\,\,\,\,\,\,\,\,\,\,\,\,$\\
\tableline$t=t_s<t_f$ & $\alpha>3$$\,\,\,\,\,\,$ &
Yes$\,\,\,\,\,\,\,\,\,\,\,\,\,\,\,\,\,\,\,\,\,\,\,\,\,\,\,\,\,\,\,\,\,\,\,$
& Occurs at $t=t_s$ & None
$\,\,\,\,\,\,\,\,\,\,\,\,\,\,\,\,\,\,\,\,\,\,\,\,\,$& Small Effect
$\,\,\,\,\,\,\,\,\,\,\,\,\,\,\,\,\,\,\,\,\,$\\
\tableline$t=t_s=t_f$ & $2<\alpha<3$ &
Yes$\,\,\,\,\,\,\,\,\,\,\,\,\,\,\,\,\,\,\,\,\,\,\,\,\,\,\,\,\,\,\,\,\,\,\,$
& Occurs at $t=t_f$ & $\epsilon_4$
$\,\,\,\,\,\,\,\,\,\,\,\,\,\,\,\,\,\,\,\,\,\,\,\,\,$& No
$\,\,\,\,\,\,\,\,\,\,\,\,\,\,\,\,\,\,\,\,\,$\\
\tableline
$t=t_s=t_f$ & $1<\alpha<2$ & Instability
$\,\,\,\,\,\,\,\,\,\,\,\,\,\,\,\,\,\,\,\,\,\,\,\,\,\,$ & Occurs at $t=t_f$ &
$\epsilon_4$, $\epsilon_3$, $\eta_H$ $\,\,\,\,\,\,\,\,\,\,\,\,\,\,\,\,\,\,$&
No $\,\,\,\,\,\,\,\,\,\,\,\,\,\,\,\,\,\,\,\,\,$\\
\tableline$t=t_s=t_f$ & $\alpha>3$$\,\,\,\,\,\,$ &
Yes$\,\,\,\,\,\,\,\,\,\,\,\,\,\,\,\,\,\,\,\,\,\,\,\,\,\,\,\,\,\,\,\,\,\,\,$
& Occurs at $t=t_f$ & None
$\,\,\,\,\,\,\,\,\,\,\,\,\,\,\,\,\,\,\,\,\,\,\,\,\,$& No
$\,\,\,\,\,\,\,\,\,\,\,\,\,\,\,\,\,\,\,\,\,$\\
\tableline
\tableline
 \end{tabular}
\end{table*}

\subsection{A Comparison of the Ordinary $R^2$ Inflation Model to the Singular
$R^2$ Model}

Having thoroughly described all the implications of a singular version of
the $R^2$ inflation model, it is worth discussing what are the differences and
similarities of it when compared to the ordinary $R^2$ inflation model. Here we
discuss briefly the qualitative and quantitative differences of the two
models. The most important difference is that in the case of the singular
$R^2$ model, inflation ends more abruptly, and also it ends at the time that
a singularity occurs. This means that it can end earlier from the ordinary
$R^2$ inflation model. Note that in the ordinary $R^2$ inflation model, inflation
ends when the first Hubble slow-roll parameter becomes of order one,
however, in the singular inflation model, inflation ends when higher order
Hubble flow parameters become singular at the Type IV singularity. This
means that the perturbative slow-roll expansion breaks at higher order, so
inflation actually ends before the first Hubble slow-roll index becomes of
order one.

Among the other differences is that inflation can end earlier in the case of
the singular inflation model. In this way, the Type IV singularity might leave its imprint
on observational quantities. However, it is possible that both the ordinary
and singular inflation models end at the same time, with the difference being
that the singular inflation model ends more abruptly.

An advantage of the singular inflation model is that it can also provide a
late-time acceleration description, without altering the theoretical
framework. Therefore, we can describe early and late-time acceleration with
the same model. The details for the various features of the singular
inflation, can be found in Table \ref{TableI}.

 A final comment is in order: The singular inflation model is approximately
described by the $F(R)$ gravity of Eq. (\ref{starobfr}) only near the
singularity at $t\simeq t_s$, but for other cosmic times, someone should use
the reconstruction technique we used in section III and find the general
behavior of the $F(R)$ gravity that generates the cosmic evolution with
Hubble rate (\ref{singstarobhub}). However, it is a formidable task to find
the general solution for the Hubble rate (\ref{singstarobhub}) since the
resulting differential equations are difficult to solve analytically. Of
course, although the $F(R)$ gravity that describes the singular inflation and
the ordinary $R^2$ inflation model are near the singularity similar, the
difference appears when the dynamics are considered, and particularly, when
the Hubble flow parameters or the Hubble slow-roll parameters are
calculated. At this level, the singularity has the effects we described in
this section.

In addition, note that the $F(R)$ gravity of the toy model near the Type IV
singularity given in Eq. (\ref{finalfrgravity}) looks similar to Eq.
(\ref{singstarobhub}), but it is totally different, since the coefficients
are negative and also due to the presence of $a_0$, which effectively acts
as a cosmological constant. In principle, $F(R)$ gravities of the form
(\ref{finalfrgravity}) can generate various cosmological scenarios, even
bounce cosmologies, see for example \cite{sergbam08bounce}. Finally, we need
to stress that the different scenarios we described in this paper are all
described by the $F(R)$ gravity appearing in Eq. (\ref{starobfr}) near the
singularity. What differentiates one from another is the time that the
singularity occurs, and the value of $\alpha$. In effect, these differ not
at the quantitative of the Hubble rate only, but at the level of the Hubble
flow and of the Hubble slow-roll parameters. Essentially, the values of
$\alpha$ determine if these are different, since some parameters that are
not divergent in some scenarios, might become divergent in another scenario.
Qualitatively these scenarios are generated by the same approximate $F(R)$
gravity, but the dynamical evolution of each scenario is different.

\subsection{Alternative Evolution Scenarios}

The singular cosmological evolutions of Eqs. (\ref{hublawsing}) and
(\ref{singstarobhub}) are not the only types of singular evolution that can
be realized. In principle, the model of Eq. (\ref{hublawsing}) can be
generalized as follows,
\begin{equation}
\label{hublawsingmodified}
H(t)=c_0+f(t)\left( t-t_s \right)^{\alpha}\, ,
\end{equation}
with $f(t)$ some smooth differentiable function, with $f(t_s)\neq 0$. For
the model (\ref{hublawsingmodified}), the early time evolution picture is
not very much affected, since the Hubble flow parameters near the
singularity at $t=t_s$ behave as follows,
\begin{align}\label{nearthesingbehavior}
& \epsilon_3\simeq -\frac{(t-t_s)^{-2+\alpha } \alpha
f(t)}{\left(c_0+\frac{4 c_0^3}{M^2}\right) M^2}+\frac{(t-t_s)^{-2+\alpha }
\alpha ^2 f(t)}{\left(c_0+\frac{4 c_0^3}{M^2}\right)
M^2}+\frac{(t-t_s)^{\alpha } \ddot{f}(t)}{\left(c_0+\frac{4
c_0^3}{M^2}\right) M^2} \\ \notag &
\epsilon_4\simeq -\frac{16}{-1+\alpha }+\frac{2}{c_0 (t-t_s) (-1+\alpha
)}+\frac{16 \alpha }{-1+\alpha }\\ \notag & -\frac{3 \alpha }{c_0 (t-t_s)
(-1+\alpha )}+\frac{\alpha ^2}{c_0 (t-t_s) (-1+\alpha )}-\frac{3
\dot{f}(t)}{c_0 (-1+\alpha ) f(t)}+\frac{3 \alpha  \dot{f}(t)}{c_0
(-1+\alpha ) f(t)}\, ,
\end{align}
however the late-time behavior of the model is very much affected, because
the singularity term controls the late-time behavior. In addition, the
singular inflation model can also be generalized accordingly, in the following
way,
\begin{equation}\label{staronsingulmodified}
H(t)\simeq H_i-\frac{M^2}{6}\left ( t-t_i\right )+f(t)\left (t-t_s
\right)^{\alpha} \, .
\end{equation}
where again, $f(t)$ is a smooth differentiable function with $f(t_s)\neq 0$.
The corresponding Hubble flow parameters near the Type IV singularity behave
as follows,
\begin{align}\label{nearthesingbehaviorstarob}
& \epsilon_3\simeq \frac{-\frac{2}{3} M^2 \left(H_i+\frac{1}{6} M^2
(-t+t_i)\right)+(t-t_s)^{-2+\alpha } (-1+\alpha ) \alpha  f(t)}{M^2
\left(H_i+\frac{1}{6} M^2 (-t+t_i)\right) \left(1+\frac{2
\left(-\frac{M^2}{6}+2 \left(H_i+\frac{1}{6} M^2
(-t+t_i)\right)^2\right)}{M^2}\right)} \\ \notag &
\epsilon_4\simeq \frac{\frac{M^4}{9}+(t-t_s)^{-2+\alpha }\left[4
\left(H_i-\frac{1}{6} M^2 (t-t_i)\right) (-1+\alpha ) \alpha
f(t)+3(-1+\alpha ) \alpha  \dot{f}(t)\right]+(t-t_s)^{-3+\alpha } (-2+\alpha
) (-1+\alpha ) \alpha  f(t)}{\left(H_i-\frac{1}{6} M^2 (t-t_i)\right)
\left(-\frac{2}{3} M^2 \left(H_i-\frac{1}{6} M^2
(t-t_i)\right)+(t-t_s)^{-2+\alpha } (-1+\alpha ) \alpha  f(t)\right)}
\, .
\end{align}
It is conceivable that the late-time behavior of this model can be different
from the singular inflation model of Eq. (\ref{singstarobhub}), with interesting
phenomenological consequences, as we demonstrate in the next section.

\section{Late-time Cosmological Evolution}

In this section we study the late-time behavior of the cosmological model of
Eq. (\ref{hublawsing}) and also of the singular inflation model
(\ref{singstarobhub}). In addition, we shall investigate how to obtain the
$\Lambda$ Cold Dark Matter ($LCDM$) model at late-times.

The cosmological evolution described by the Hubble rate of Eq.
(\ref{hublawsing}), has some appealing phenomenological features, since it
can describe late-time and early-time acceleration, and thus it offers the
possibility of a unified description of these acceleration eras. In order to
see this, we shall calculate the EoS $w_{eff}$ corresponding to the Hubble
rate (\ref{hublawsing}), and focus on the behavior of the EoS near the Type
IV singularity, which corresponds to early times. Additionally we shall
examine the behavior of the EoS at late times. The EoS for an $F(R)$ gravity
is given by \cite{reviews1},
\begin{equation}\label{eosdeded}
w_{eff}=-1-\frac{2\dot{H}}{3H^2}\, ,
\end{equation}
so by substituting the Hubble rate (\ref{hublawsing}), the EoS reads,
\begin{equation}\label{gfgfgdss}
w_{eff}=-1-\frac{2 f_0 (t-t_s)^{-1+\alpha } \alpha }{3 \left(c_0+f_0
(t-t_s)^{\alpha }\right)^2}\, .
\end{equation}
So at early times, since the term $c_0$ dominates the denominator, the EoS
becomes,
\begin{equation}\label{gfgfgdss12}
w_{eff}= -1-\frac{2 f_0 (t-t_s)^{-1+\alpha } \alpha }{3 c_0^2}\, .
\end{equation}
Since $c_0\gg f_0$, this EoS describes a nearly de Sitter evolution,
which at $t=t_s$, where the Type IV singularity occurs, it exactly describes
a de Sitter evolution, since the term $\sim (t-t_s)^{-1+\alpha } $, becomes
equal to zero. Now we turn our focus at late times, which means that the
cosmic time is larger or equal to present times $t_p\sim 10^{17}$sec. At
late times, the EoS becomes,
\begin{equation}\label{eoslatetims}
w_{eff}\simeq -1-\frac{2 t^{-1-\alpha } \alpha }{3 f_0}\, ,
\end{equation}
and since the second term is very small for $t\geq t_p$, the EoS of Eq.
(\ref{eoslatetims}) describes a nearly de Sitter evolution. Notice that the
evolution is slightly phantom, a feature quite interesting, since present
time observations favor a nearly phantom evolution \cite{phantom}. Hence as
we demonstrated, both early and late-time acceleration can be described by
the same cosmological model appearing in Eq. (\ref{hublawsing}).

With regards to the singular inflation model (\ref{singstarobhub}), the EoS at
early times is equal to,
\begin{equation}\label{earlytimesstarob}
w_{eff}\simeq -1+\frac{M^2}{9 \left(H_i-\frac{1}{6} M^2 (t-t_i)\right)^2}\,
,
\end{equation}
which clearly describes quintessential acceleration. Note that the EoS of
Eq. (\ref{earlytimesstarob}), is identical to the EoS corresponding to the
ordinary $R^2$ inflation model. At late-times, the EoS of the singular inflation
model behaves as follows,
\begin{equation}\label{latetimesingstarob}
w_{eff}=-1-\frac{2 t^{-1-\alpha } \alpha }{3 f_0}\, ,
\end{equation}
which since $t\simeq 10^{17}$sec, it describes nearly de Sitter
acceleration, because the term $\sim t^{-1-\alpha }$ is relatively small.
Therefore, what we have as a resulting picture is quite appealing, since the
Type IV singularity in the $R^2$ inflation Hubble rate offers the possibility to
simultaneously describe early and late-time acceleration with a single
model. In addition, at early times, the effect of the singularity is to make
the dynamical evolution unstable at the singularity, while at late-times it
is possible to have nearly de Sitter evolution. Here we need to clarify an important issue before we continue: The singular inflation Hubble rate is not by any means generated by an $F(R)$ gravity of the form $F(R)\sim R+\beta R^2$. What we proved in the previous sections is that the $F(R)$ gravity that can generate the singular evolution (\ref{singstarobhub}), is approximately an $R^2$ gravity only near the Type IV singularity. The general form of the $F(R)$ gravity that generates the Hubble rate (\ref{singstarobhub}), is rather difficult to find analytically, but we can infer its probable behavior from the EoS study we performed in this section. Since it seems that both at early and late times, the evolution is nearly de Sitter, this means that the $F(R)$ gravity which generates the Hubble rate (\ref{singstarobhub}) obviously leads to two de Sitter vacua, one occurring at early times and the other occurring at late times. So we could say that the $F(R)$ gravity which generates the Hubble rate (\ref{singstarobhub}), at early times behaves as $\sim R^2$ and at late times, where the curvature is small, it behaves as a small constant $\sim \Lambda $. An $F(R)$ model that behaves approximately as we described is the following (the last term describes the dark energy sector \cite{sergeinewplb}),
\begin{equation}\label{sergeinewfr}
F(R)=R+\frac{1}{6M^2}R^2+\frac{f_0\Lambda_i}{f_0+\frac{1}{\Lambda_i}\left((R-R_0)^{2n+1}+R_0^{2n+1}\right)}\, ,
\end{equation}
with $n=1,2,...$, which at early times behaves as $F(R)\sim R^2$ and at late times as $F(R)\sim \Lambda_i$, so effectively it has two de Sitter vacua. However, the detailed form of the $F(R)$ gravity that exactly generates the Hubble rate (\ref{singstarobhub}) is difficult to find, but judging from its qualitative features, it should behave as that of Eq. (\ref{sergeinewfr}). 

In Table \ref{TableII} we
gathered the results for the models of Eqs. (\ref{hublawsing}) and
(\ref{singstarobhub}), with regards to the EoS at late times and early
times.
\begin{table*}[h]
\small
\caption{\label{TableII}The EoS for the cosmological models of Eqs.
(\ref{hublawsing}) and (\ref{singstarobhub})}
\begin{tabular}{@{}crrrrrrrrrrr@{}}
\tableline
\tableline
\tableline
Cosmological Model & EoS at late-time  & EoS at early-time
\\\tableline
$H(t)=c_0+f_0\left( t-t_s \right)^{\alpha}$ &  Nearly de Sitter & Nearly de
Sitter\\
\tableline
$H(t)\simeq H_i-\frac{M^2}{6}\left ( t-t_i\right )+f_0\left (t-t_s
\right)^{\alpha}$ & Nearly de Sitter &
$\,\,\,\,\,\,\,\,\,\,\,\,\,\,$Quintessential Acceleration \\
\tableline
\tableline
\tableline
 \end{tabular}
\end{table*}
As we discussed in the previous section, the Type IV singular cosmological
evolutions described by Eqs. (\ref{hublawsing}) and (\ref{singstarobhub}),
have the most simple forms, and generalizations appear in Eqs.
(\ref{staronsingulmodified}) and (\ref{hublawsingmodified}). In the rest of
this section we shall be interested in the generalized cosmological
evolutions (\ref{staronsingulmodified}) and (\ref{hublawsingmodified}), and
we shall investigate how the function $f(t)$ appearing in both Eqs.
(\ref{staronsingulmodified}) and (\ref{hublawsingmodified}) should behave in
order to have the $LCDM$ model at late-times.

Before we proceed to the $LCDM$ production at late-times, let us investigate
what is the effect of the function $f(t)$ on the corresponding EoS at both
early and late times. We start off with the model
(\ref{hublawsingmodified}), and the corresponding EoS becomes,
\begin{equation}\label{modeosgener}
w_{eff}=-1-\frac{2 \left((t-t_s)^{-1+\alpha } \alpha  f(t)+(t-t_s)^{\alpha }
\dot{f}(t)\right)}{3 \left(c_0+(t-t_s)^{\alpha } f(t)\right)^2}\, ,
\end{equation}
so at early times, near the Type IV singularity, this becomes,
\begin{equation}\label{modeosgenerearly}
w_{eff}\simeq -1-\frac{2 \left((t-t_s)^{-1+\alpha } \alpha
f(t)+(t-t_s)^{\alpha } \dot{f}(t)\right)}{3 c_0^2}\, ,
\end{equation}
while at late times, the EoS becomes,
\begin{equation}\label{modeosgenerearlyelate}
w_{eff}\simeq -1-\frac{2 t^{-1-\alpha } \alpha }{3 f(t)}-\frac{2 t^{-\alpha
} \dot{f}(t)}{3 f(t)^2}\, .
\end{equation}
From Eqs. (\ref{modeosgenerearly}) and (\ref{modeosgenerearlyelate}) it is
obvious that both the early and late-time behavior of the EoS are affected
by the presence of the $f(t)$. For the singular inflation Hubble rate, the EoS
becomes,
\begin{equation}\label{modeosgenerearlystarob}
w_{eff}\simeq -1-\frac{2 \left(-\frac{M^2}{6}+(t-t_s)^{-1+\alpha } \alpha
f(t)+(t-t_s)^{\alpha } \dot{f}(t)\right)}{3 \left(H_i-\frac{1}{6} M^2
(t-t_i)+(t-t_s)^{\alpha } f(t)\right)^2}\, ,
\end{equation}
which at early times becomes,
\begin{equation}\label{modeosgenerearlyelatestarob}
w_{eff}\simeq -1+\frac{M^2}{9 \left(H_i-\frac{1}{6} M^2 (t-t_i)\right)^2}\,
,
\end{equation}
while at late times the singular inflation EoS becomes,
\begin{equation}\label{modeosgenerearlyelatestaroblate}
w_{eff}\simeq -1-\frac{2 t^{-1-\alpha } \alpha }{3 f(t)}-\frac{2 t^{-\alpha
} \dot{f}(t)}{3 f(t)^2}\, .
\end{equation}
It is obvious that for the singular inflation model, only the late-time behavior
of the EoS is affected by the function $f(t)$, a feature that renders the
singular inflation model quite appealing.

In the following two sections we shall investigate what the function $f(t)$
should be so the late-time cosmological evolution is described by the $LCDM$
Hubble rate. Having found the function $f(t)$, we shall investigate what is
its effect on the EoS.

\subsection{Late-time $LCDM$ Evolution from $H(t)=c_0+f(t)\left( t-t_s
\right)^{\alpha}$}

In this section we shall investigate how should $f(t)$ behave so that the
late-time behavior of the Hubble rate (\ref{hublawsingmodified}) resembles
the one corresponding to the $LCDM$ model, with the latter being equal to,
\begin{equation}\label{lcdmhub}
H(t)=\sqrt{H_0^2+\frac{\kappa^2\rho_0}{3}a(t)^{-3}}\, ,
\end{equation}
with $a(t)$ the scale factor of the $LCDM$ model. Then, in order the model
(\ref{hublawsingmodified}) becomes approximately the $LCDM$ model at late
times, after some simple calculations, we end up that the $f(t)$ should be
of the form,
\begin{equation}\label{gflatesingularsimplemodel}
f(t)\simeq \frac{e^{2 H_0 t} t^{-\alpha }}{2^{2/3}
\left(H_0^2\right)^{2/3}}\, .
\end{equation}
By using this, and substituting in Eq. (\ref{modeosgenerearly}), we can
easily find how the EoS is affected by the form of the $f(t)$ function,
which becomes approximately a de Sitter evolution, $w_{eff}\simeq -1$.
Correspondingly, at late times, the EoS behaves as,
\begin{equation}\label{wesoslatetimesforsecificf}
w_{eff}\simeq -1-\frac{4}{3} 2^{2/3} e^{-2 H_0 t} H_0
\left(H_0^2\right)^{2/3}\, ,
\end{equation}
which describes a nearly de Sitter evolution. Therefore, in the context of
the model (\ref{hublawsingmodified}) it is possible to describe early and
late-time evolution, with the late-time evolution being similar to the
$LCDM$ model.

\subsection{ Late-time $LCDM$ Evolution from the Modified Singular Inflation
Model}

Now let us see what is the effect of the $f(t)$ function given in Eq.
(\ref{gflatesingularsimplemodel}) on the EoS of the modified singular inflation
model (\ref{staronsingulmodified}). Substituting by using $f(t)$, the
early-time behavior (near the Type IV singularity) of the EoS becomes,
\begin{equation}\label{earlyeossingumodstaro}
w_{eff}\simeq -1+\frac{M^2}{9 \left(H_i-\frac{1}{6} M^2 (t-t_i)\right)^2}\,
,
\end{equation}
which is identical to Eq. (\ref{earlytimesstarob}), so practically the form
of the function $f(t)$ given in Eq. (\ref{gflatesingularsimplemodel}) leaves
unaffected the modified singular inflation model at early times. Correspondingly
at late times, the EoS behaves as,
\begin{equation}\label{latetimemodsingstarobmod}
w_{eff}\simeq -1-\frac{4}{3} 2^{2/3} e^{-2 H_0 t} H_0^{7/3}\, ,
\end{equation}
which describes a nearly de Sitter evolution. In conclusion, the present
case is particularly interesting, since we can describe early and late-time
acceleration with the modified singular inflation model
(\ref{staronsingulmodified}), with the late-time one being similar to the
$LCDM$ model. It is important to note that the early-time behavior of the
model is not affected if the $f(t)$ function is given by Eq.
(\ref{gflatesingularsimplemodel}), since the Hubble rate at early times
becomes,
\begin{equation}\label{earlytimehubmodsingstarobmodel}
H(t)\simeq H_i-\frac{1}{6} M^2 (t-t_i)+\frac{e^{2 H_0 t} t^{-\alpha }
(t-t_s)^{\alpha }}{2^{2/3} \left(H_0^2\right)^{2/3}}\,.
\end{equation}
The difference with the singular inflation model of Eq. (\ref{singstarobhub}),
is the presence of the last term $\sim \frac{e^{2 H_0 t} t^{-\alpha }
(t-t_s)^{\alpha }}{2^{2/3} \left(H_0^2\right)^{2/3}}$, which for early times
($t\sim 10^{-35}$sec) it becomes $\sim \frac{ t^{-\alpha } (t-t_s)^{\alpha
}}{2^{2/3} \left(H_0^2\right)^{2/3}}$, which is relatively small compared to
the other terms\footnote{The exponential term in Eq.
(\ref{earlytimehubmodsingstarobmodel}) becomes nearly equal to one, for
early times.}. The final qualitative picture is quite interesting, since the
vacuum modified singular inflation model of Eq. (\ref{singstarobhub}), at early
times has the features of the ordinary $R^2$ inflation model, and at late-times
it can describe a Universe with a nearly $LCDM$ evolution. Of course, there
exist many other choices for the function $f(t)$, that could lead to
alternative scenarios at late times, but we omit this study for brevity. We need to note that, as it is obvious, the late-time behavior of the $F(R)$ gravity is not given anymore from Eq. (\ref{starobfr}). In relation to that, a final comment is in order. In the literature there exist viable promising $F(R)$ gravity models, which are consistent with solar system tests and also
generate a $LCDM$ late-time evolution, like the exponential $F(R)$ model
studied in Ref. \cite{sergeiexp}, in which case the $F(R)$ gravity has the
form,
\begin{equation}\label{expformfr}
F(R)=R-2\Lambda_{eff}\left( e^{-bR}-1\right)\, ,
\end{equation}
which when $R\gg b$, it becomes $F(R)\sim R+2 \Lambda_{eff}$. So in the
presence of matter, the $F(R)$ gravity of Eq. (\ref{expformfr}) can yield a
$LCDM$-like behavior.

\section{Discussion and Conclusions}

In this paper we studied some crucial aspects and implications of singular
inflation in the Jordan frame, in the context of vacuum $F(R)$ gravity. We
used a toy inflationary model and also a singular version of the $R^2$ inflation. In all cases, the singularity during the inflationary era was
assumed to be a Type IV singularity, which is a ``mild'' singularity
phenomenologically, meaning that the observable quantities that can be
defined at the three dimensional spacelike hypersurface corresponding to the
time instance that the singularity occurs, are finite. However, the Type IV
singularity affects significantly the parameters that control the
inflationary dynamics, since these are rendered singular at the time
instance that the singularity occurs. We studied two kinds of dynamical
parameters, the Hubble flow parameters and also the Hubble slow-roll
parameters, and as we evinced some of these are singular at the Type IV
singularity. With regards to the Hubble flow parameters, these are strongly
affected by the form of the $F(R)$ gravity, and this might alter the
singularity structure of these Hubble flow parameters. The presence of
singularities in the parameters that determine the dynamics of the
cosmological system, clearly indicates that the dynamical evolution of the
system becomes abruptly interrupted and the system becomes dynamically
unstable. This means that the cosmological solution that described the
evolution of the system up to that point, ceases to describe the system and
therefore cannot be a final attractor of the cosmological dynamical system.
Clearly, the presence of singularities in the Hubble flow parameters could
be an indicator for a graceful exit from inflation, as we claimed. Note
however that the presence of singularities is not a mechanism for graceful
exit from inflation, but indicates that there is probably some mechanism
that generates graceful exit at the point that the singularities occur.
Indeed, as we demonstrated in the case of the toy model, the $F(R)$
cosmological system we studied has an unstable de Sitter point, which is
actually a solution to the system at the time instance that the singularity
occurs. The de Sitter point was shown to be unstable towards linear
perturbations, so in effect, curvature perturbations will generate graceful
exit from inflation. For a thorough study on this kind of graceful exit from
inflation mechanism, we refer to \cite{sergeitraceanomaly}. In principle,
there can also be other mechanisms that can generate graceful exit from
inflation, like for example tachyonic instabilities in scalar-tensor
theories \cite{encyclopedia}, or even higher loop quantum effects which can
also generate a sufficient amount of curvature fluctuations that can end
inflation. Finally, we discussed the late-time dynamics of the cosmological
evolution we used in this paper. As we evinced, the toy model can also
describe late-time acceleration, so it can describe successfully
simultaneously early and late-time acceleration. For the case of the
singular inflation model, the results are particularly interesting, since there
are various scenarios that lead to results compatible with the observations.
Instabilities occur at higher order parameters of the slow-roll expansion,
and as we claimed, this might be an indication of graceful exit occurring at
higher order in slow-roll expansion. In addition, in most cases, the effect
of the singularity is to modify the dynamics of inflation, but in some cases
it may have observable effects. Actually the Type IV
singularity modifies the spectral index of primordial curvature
perturbations, but as we showed the difference it causes is significantly small.

A direct comparison of the singular inflation model to the ordinary $R^2$ inflation
model, indicates that in the singular inflation model, inflation might end
earlier than the ordinary $R^2$ inflation model, and in a more abrupt way. Also
it is possible that both the singular and ordinary $R^2$ inflation model end at
the same time, but in the case of the singular inflation model, inflation ends
more abruptly again.

An interesting issue that we did not address in this paper, is to
investigate the effects of Loop Quantum Cosmology \cite{LQC} corrections on
the inflationary dynamics of $F(R)$ gravities in the Jordan frame, in the
presence of a Type IV singularity. In addition to Loop Quantum Cosmology
corrections, quantum effects should also be taken into account, since these
are known to render the finite time singularities milder
\cite{Nojiri:2005sx}. Apart from Loop Quantum Cosmology effects, the effects of ordinary matter
perfect fluids should be examined too. We hope to address these issues in
detail in a future publication.

It is interesting to note that the effects of a Type IV singularity in a
bouncing cosmology are quite more severe than in the inflationary solution
we described here. Actually, as probably expected, the inflationary
description does not apply, but also singularities might appear in the
dynamical quantities that determine the inflationary evolution, for
example the comoving Hubble radius. We hope to address this issue in a
future work.  Finally, an interesting question is to see if the Type IV
singularity occurs in the Einstein frame, if we perform a conformal
transformation. However, a consistent study of this issue would require the
full analytic behavior of the $F(R)$ gravity for the singular inflationary
models we studied in this paper, so we refer from going into details on
this, since our solution is just an approximation near the Type IV
singularity.

\section*{Acknowledgments}

This work is supported by MINECO (Spain), project
 FIS2013-44881 (S.D.O) and by Min. of Education and Science of Russia (S.D.O
and V.K.O).

\section*{Appendix: Detailed Form of the Parameters Appearing in the Main
Text}

In this Appendix we provide the detailed form of various parameters
appearing in the main text of the paper. We start of with the parameters
$A$, $B$ and $C$, which appear in Eq. (\ref{finalxr}), the detailed form of
which is,
\begin{equation}\label{dfsgdffdfdfdf123}
A=\frac{c_1 c_0}{1+c_0},\,\,\,B=-\frac{12 c_0^3c_1}{1+c_0},\,\,\,C=-\frac{12
c_0^4c_1}{1+c_0}\, .
\end{equation}
Also, the parameters $a_0$ and $a_2$ appearing in Eq. (\ref{finalfrgravity})
are equal to,
\begin{equation}\label{aoa2}
 a_0=-\frac{15 c_0^2}{2}-\frac{6 c_0^2 c_2}{1+c_0}-\frac{6 c_0^3
c_2}{1+c_0},\,\,\,a_2=-\frac{1}{32 c_0^2}\, .
\end{equation}
In addition, the full analytic form of the Hubble flow parameter
$\epsilon_4$, calculated for $F(R)$ gravity appearing in Eq.
(\ref{finalfrgravity}), is equal to,
\begin{align}\label{r2epsilon4}
& \epsilon_4=\frac{2 f_0 (t-t_s)^{-3+\alpha } \alpha }{\left(c_0+f_0
(t-t_s)^{\alpha }\right) \left(4 f_0 \left(c_0+f_0 (t-t_s)^{\alpha }\right)
(t-t_s)^{-1+\alpha } \alpha +f_0 (t-t_s)^{-2+\alpha } (-1+\alpha ) \alpha
\right)}\\ \notag & -\frac{4 c_0 f_0 (t-t_s)^{-2+\alpha } \alpha
}{\left(c_0+f_0 (t-t_s)^{\alpha }\right) \left(4 f_0 \left(c_0+f_0
(t-t_s)^{\alpha }\right) (t-t_s)^{-1+\alpha } \alpha +f_0 (t-t_s)^{-2+\alpha
} (-1+\alpha ) \alpha \right)}\\ \notag &
-\frac{4 f_0^2 (t-t_s)^{-2+2 \alpha } \alpha }{\left(c_0+f_0 (t-t_s)^{\alpha
}\right) \left(4 f_0 \left(c_0+f_0 (t-t_s)^{\alpha }\right)
(t-t_s)^{-1+\alpha } \alpha +f_0 (t-t_s)^{-2+\alpha } (-1+\alpha ) \alpha
\right)}\\ \notag &-\frac{3 f_0 (t-t_s)^{-3+\alpha } \alpha
^2}{\left(c_0+f_0 (t-t_s)^{\alpha }\right) \left(4 f_0 \left(c_0+f_0
(t-t_s)^{\alpha }\right) (t-t_s)^{-1+\alpha } \alpha +f_0 (t-t_s)^{-2+\alpha
} (-1+\alpha ) \alpha \right)}\\ \notag & +\frac{4 c_0 f_0
(t-t_s)^{-2+\alpha } \alpha ^2}{\left(c_0+f_0 (t-t_s)^{\alpha }\right)
\left(4 f_0 \left(c_0+f_0 (t-t_s)^{\alpha }\right) (t-t_s)^{-1+\alpha }
\alpha +f_0 (t-t_s)^{-2+\alpha } (-1+\alpha ) \alpha \right)}\\ \notag &
+\frac{8 f_0^2 (t-t_s)^{-2+2 \alpha } \alpha ^2}{\left(c_0+f_0
(t-t_s)^{\alpha }\right) \left(4 f_0 \left(c_0+f_0 (t-t_s)^{\alpha }\right)
(t-t_s)^{-1+\alpha } \alpha +f_0 (t-t_s)^{-2+\alpha } (-1+\alpha ) \alpha
\right)}\\ \notag & +\frac{f_0 (t-t_s)^{-3+\alpha } \alpha ^3}{\left(c_0+f_0
(t-t_s)^{\alpha }\right) \left(4 f_0 \left(c_0+f_0 (t-t_s)^{\alpha }\right)
(t-t_s)^{-1+\alpha } \alpha +f_0 (t-t_s)^{-2+\alpha } (-1+\alpha ) \alpha
\right)}\, .
\end{align}
Finally, the same Hubble flow parameter, when calculated for a general
$F(R)$ gravity, reads,
\begin{align}\label{r2epsilon4}
& \epsilon_4=\frac{6 f_0 F_{RRR} (t-t_s)^{\alpha } \alpha  \left(-1+4 c_0
(t-t_s)+4 f_0 (t-t_s)^{1+\alpha }+\alpha \right)^2}{F_{RR} \left(c_0+f_0
(t-t_s)^{\alpha }\right) (t-t_s)^2 \left(-1+4 c_0 (t-t_s)+4 f_0
(t-t_s)^{1+\alpha }+\alpha \right)}\\ \notag &
+\frac{F_{RR} (t-t_s) \left(2+4 c_0 (t-t_s) (-1+\alpha )-3 \alpha +\alpha
^2+4 f_0 (t-t_s)^{1+\alpha } (-1+2 \alpha )\right)}{F_{RR} \left(c_0+f_0
(t-t_s)^{\alpha }\right) (t-t_s)^2 \left(-1+4 c_0 (t-t_s)+4 f_0
(t-t_s)^{1+\alpha }+\alpha \right)}
\, .
\end{align}


\newpage

\begin{thebibliography}{99}


\bibitem{hawkingpenrose}
S.~W.~Hawking and R.~Penrose,
Proc.\ Roy.\ Soc.\ Lond.\ A {\bf 314} (1970) 529.




\bibitem{Virbhadra:2002ju}
  K.~S.~Virbhadra and G.~F.~R.~Ellis,
  Phys.\ Rev.\ D {\bf 65} (2002) 103004.



\bibitem{Nojiri:2005sx}
S.~Nojiri, S.~D.~Odintsov and S.~Tsujikawa,
Phys.\ Rev.\ D {\bf 71}, 063004 (2005) [arXiv:hep-th/0501025].






\bibitem{ref5}
R.~R.~Caldwell, M.~Kamionkowski and N.~N.~Weinberg,
Phys.\ Rev.\ Lett.\ {\bf 91}, 071301 (2003)
[arXiv:astro-ph/0302506].;\\
B.~McInnes,
JHEP {\bf 0208} (2002) 029
[arXiv:hep-th/0112066]; \\
S.~Nojiri and S.~D.~Odintsov,
Phys.\ Lett.\ B {\bf 562}, 147 (2003)
[arXiv:hep-th/0303117]; \\
S.~Nojiri and S.~D.~Odintsov,
Phys.\ Rev.\ D {\bf 72} (2005) 023003
[hep-th/0505215]; Phys.\ Rev.\ D {\bf 70} (2004) 103522
  [hep-th/0408170].;\\
V.~Gorini, A.~Kamenshchik and U.~Moschella,
Phys.\ Rev.\ D {\bf 67} (2003) 063509
[astro-ph/0209395]; \\
E.~Elizalde, S.~Nojiri and S.~D.~Odintsov,
  Phys.\ Rev.\ D {\bf 70} (2004) 043539
  [hep-th/0405034]. ; \\
V.~Faraoni,
Int.\ J.\ Mod.\ Phys.\ D {\bf 11}, 471 (2002)
[arXiv:astro-ph/0110067]; \\
P.~Singh, M.~Sami and N.~Dadhich,
Phys.\ Rev.\ D {\bf 68}, 023522 (2003)
[arXiv:hep-th/0305110]; \\
C.~Csaki, N.~Kaloper and J.~Terning,
Annals Phys.\ {\bf 317}, 410 (2005)
[arXiv:astro-ph/0409596]; \\
P.~X.~Wu and H.~W.~Yu,
Nucl.\ Phys.\ B {\bf 727}, 355 (2005)
[arXiv:astro-ph/0407424]; \\
S.~Nesseris and L.~Perivolaropoulos,
Phys.\ Rev.\ D {\bf 70}, 123529 (2004)
[arXiv:astro-ph/0410309]; \\
M.~Sami and A.~Toporensky,
Mod.\ Phys.\ Lett.\ A {\bf 19}, 1509 (2004)
[arXiv:gr-qc/0312009]; \\
H.~Stefancic,
Phys.\ Lett.\ B {\bf 586}, 5 (2004)
[arXiv:astro-ph/0310904]; \\
L.~P.~Chimento and R.~Lazkoz,
Phys.\ Rev.\ Lett.\ {\bf 91}, 211301 (2003)
[arXiv:gr-qc/0307111]; \\
Mod.\ Phys.\ Lett.\ A {\bf 19}, 2479 (2004)
[arXiv:gr-qc/0405020]; \\
J.~G.~Hao and X.~Z.~Li,
Phys.\ Lett.\ B {\bf 606}, 7 (2005)
[arXiv:astro-ph/0404154]; \\
E.~Babichev, V.~Dokuchaev and Yu.~Eroshenko,
Class.\ Quant.\ Grav.\ {\bf 22}, 143 (2005)
[arXiv:astro-ph/0407190]; \\
X.~F.~Zhang, H.~Li, Y.~S.~Piao and X.~M.~Zhang,
Mod.\ Phys.\ Lett.\ A {\bf 21}, 231 (2006)
[arXiv:astro-ph/0501652]; \\
M.~P.~Dabrowski and T.~Stachowiak,
Annals Phys.\ {\bf 321}, 771 (2006)
[arXiv:hep-th/0411199]; \\
F.~S.~N.~Lobo,
Phys.\ Rev.\ D {\bf 71}, 084011 (2005)
[arXiv:gr-qc/0502099]; \\
R.~G.~Cai, H.~S.~Zhang and A.~Wang,
Commun.\ Theor.\ Phys.\ {\bf 44}, 948 (2005)
[arXiv:hep-th/0505186]; \\
I.~Y.~Aref'eva, A.~S.~Koshelev and S.~Y.~Vernov,
Theor.\ Math.\ Phys.\ {\bf 148}, 895 (2006) [Teor.\ Mat.\ Fiz.\ {\bf
148}, 23 (2006)]
[arXiv:astro-ph/0412619]; \\
Phys.\ Rev.\ D {\bf 72}, 064017 (2005)
[arXiv:astro-ph/0507067]; \\
H.~Q.~Lu, Z.~G.~Huang and W.~Fang,
arXiv:hep-th/0504038; \\
W.~Godlowski and M.~Szydlowski,
Phys.\ Lett.\ B {\bf 623}, 10 (2005)
[arXiv:astro-ph/0507322]; \\
J.~Sola and H.~Stefancic,
Phys.\ Lett.\ B {\bf 624}, 147 (2005)
[arXiv:astro-ph/0505133]; \\
B.~Guberina, R.~Horvat and H.~Nikolic,
Phys.\ Rev.\ D {\bf 72}, 125011 (2005)
[arXiv:astro-ph/0507666]; \\
M.~P.~Dabrowski, C.~Kiefer and B.~Sandhofer,
Phys.\ Rev.\ D {\bf 74}, 044022 (2006)
[arXiv:hep-th/0605229]



\bibitem{Barrow:2015ora}
J.~D.~Barrow and A.~A.~H.~Graham,
  Phys.\ Rev.\ D {\bf 91}, no. 8, 083513 (2015)
  [arXiv:1501.04090 [gr-qc]].


\bibitem{noo1}
  S.~Nojiri, S.~D.~Odintsov and V.~K.~Oikonomou,
  Phys.\ Rev.\ D {\bf 91} (2015) 8,  084059
  [arXiv:1502.07005 [gr-qc]]; Phys.\ Lett.\ B {\bf 747} (2015) 310
  [arXiv:1506.03307]


\bibitem{noo2}

S.~D.~Odintsov and V.~K.~Oikonomou,
  arXiv:1504.01772;  Phys.\ Rev.\ D {\bf 92} (2015) 2,  024058
  [arXiv:1507.05273  ]; Phys.\ Rev.\ D {\bf 92} (2015) 2,  024016
  [arXiv:1504.06866  ]


  \bibitem{noo6} S.~Nojiri, S.~D.~Odintsov, V.~K.~Oikonomou and
E.~N.~Saridakis,
  JCAP {\bf 1509} (2015) 09,  044
  [arXiv:1503.08443 [gr-qc]]




\bibitem{Barrow:2004xh}
J.~D.~Barrow,
Class.\ Quant.\ Grav.\  {\bf 21} (2004) L79 [gr-qc/0403084]. ;
\bibitem{Barrow:2004hk}
J.~D.~Barrow,
Class.\ Quant.\ Grav.\  {\bf 21} (2004) 5619 [gr-qc/0409062].



\bibitem{barrow}
S.~Nojiri and S.~D.~Odintsov,
  Phys.\ Lett.\ B {\bf 595} (2004) 1
  [hep-th/0405078]; \\
Z.~Keresztes, L.~\' A.~Gergely, A.~Y.~Kamenshchik, V.~Gorini and
D.~Polarski,
Phys.\ Rev.\ D {\bf 88} (2013) 023535
[arXiv:1304.6355 [gr-qc]];
M.~Bouhmadi-Lopez, C.~Kiefer, B.~Sandhofer and P.~V.~Moniz,
Phys.\ Rev.\ D {\bf 79}, 124035 (2009)
[arXiv:0905.2421 [gr-qc]];
V.~Sahni and Y.~Shtanov,
JCAP {\bf 0311}, 014 (2003)
[arXiv:astro-ph/0202346];
K.~Lake,
Class.\ Quant.\ Grav.\ {\bf 21}, L129 (2004)
[arXiv:gr-qc/0407107]; \\
J.~D.~Barrow and C.~G.~Tsagas,
Class.\ Quant.\ Grav.\ {\bf 22}, 1563 (2005)
[arXiv:gr-qc/0411045]; \\
M.~P.~Dabrowski,
Phys.\ Rev.\ D {\bf 71}, 103505 (2005)
[arXiv:gr-qc/0410033]; \\
Phys. Lett. B {\bf 625}, 184 (2005); L.~Fernandez-Jambrina and
R.~Lazkoz,
Phys.\ Rev.\ D {\bf 70}, 121503 (2004) [arXiv:gr-qc/0410124];
Phys.\ Rev.\ D {\bf 74}, 064030 (2006) [arXiv:gr-qc/0607073];
arXiv:0805.2284 [gr-qc]; \\
P.~Tretyakov, A.~Toporensky, Y.~Shtanov and V.~Sahni,
Class.\ Quant.\ Grav.\ {\bf 23}, 3259 (2006)
[arXiv:gr-qc/0510104]; \\
H.~Stefancic,
Phys.\ Rev.\ D {\bf 71}, 084024 (2005)
[arXiv:astro-ph/0411630]; \\
A.~V.~Yurov, A.~V.~Astashenok and P.~F.~Gonzalez-Diaz,
Grav.\ Cosmol.\ {\bf 14}, 205 (2008)
[arXiv:0705.4108 [astro-ph]]; \\
K.~Bamba, S.~D.~Odintsov, L.~Sebastiani and S.~Zerbini, Eur.\ Phys.\
J.\ C {\bf 67} (2010) 295
[arXiv:0911.4390]; \\
I.~Brevik and O.~Gorbunova,
Eur.\ Phys.\ J.\ C {\bf 56}, 425 (2008)
[arXiv:0806.1399 [gr-qc]]; \\
M.~Bouhmadi-Lopez, P.~F.~Gonzalez-Diaz and P.~Martin-Moruno,
Phys.\ Lett.\ B {\bf 659}, 1 (2008) [arXiv:gr-qc/0612135];
arXiv:0707.2390 [gr-qc]; \\
C.~Cattoen and M.~Visser,
Class.\ Quant.\ Grav.\ {\bf 22}, 4913 (2005)
[arXiv:gr-qc/0508045]; \\
J.~D.~Barrow and S.~Z.~W.~Lip,
arXiv:0901.1626 [gr-qc]; \\
M.~Bouhmadi-Lopez, Y.~Tavakoli and P.~V.~Moniz,
arXiv:0911.1428 [gr-qc].; \\
J.~D.~Barrow, A.~B.~Batista, J.~C.~Fabris, M.~J.~S.~Houndjo and
G.~Dito,
Phys.\ Rev.\ D {\bf 84} (2011) 123518
[arXiv:1110.1321 [gr-qc]];
M.~Bouhmadi-López, C.~Kiefer and M.~Krämer,
  Phys.\ Rev.\ D {\bf 89} (2014) 6,  064016
  [arXiv:1312.5976 [gr-qc]]; M.~Bouhmadi-Lopez, P.~Chen and Y.~W.~Liu,
  Eur.\ Phys.\ J.\ C {\bf 73} (2013) 2546
  [arXiv:1302.6249 [gr-qc]].


\bibitem{barrowslowroll}  A.~R.~Liddle, P.~Parsons and J.~D.~Barrow,
  Phys.\ Rev.\ D {\bf 50} (1994) 7222
  [astro-ph/9408015]; \\
   A.~R.~Liddle and D.~H.~Lyth,
  Phys.\ Lett.\ B {\bf 291} (1992) 391
  [astro-ph/9208007].;\\
   E.~J.~Copeland, E.~W.~Kolb, A.~R.~Liddle and J.~E.~Lidsey,
  Phys.\ Rev.\ D {\bf 48} (1993) 2529
  [hep-ph/9303288].


\bibitem{inflation}

V. F. Mukhanov, H. A. Feldman and R. H. Brandenberger, Phys. Rept. {\bf
215}, 203 (1992).
V.~Mukhanov, ``Physical foundations of cosmology,''
Cambridge, UK: Univ. Pr. (2005) 421 p;
D.~S.~Gorbunov and V.~A.~Rubakov, ``Introduction to the theory of
the early universe: Cosmological perturbations and inflationary
theory,''
Hackensack, USA: World Scientific (2011) 489 p;
%
(2005) 421 p; D. S. Gorbunov, V. A. Rubakov, Introduction to the theory of
the early Universe: Cosmological perturbations and inflationary theory,
Hackensack, USA, World Scientific (2011) 489 p
A.~Linde,
arXiv:1402.0526 [hep-th];
K.~Bamba and S.~D.~Odintsov,
  Symmetry {\bf 7} (2015) 220
  [arXiv:1503.00442 [hep-th]];
D.~H.~Lyth and A.~Riotto,
Phys.\ Rept.\  {\bf 314} (1999) 1 [hep-ph/9807278].;
 S.~P.~Miao and R.~P.~Woodard,
  JCAP {\bf 1509} (2015) 09,  022
  [arXiv:1506.07306 [astro-ph.CO]].



\bibitem{lehners}

J.~L.~Lehners,
  Phys.\ Rev.\ D {\bf 86} (2012) 043518
  [arXiv:1206.1081 [hep-th]].





\bibitem{reviews1}

S. Nojiri and S. D. Odintsov,  Phys.Rept. 505 (2011) 59 [arXiv:1011.0544];
Int.\ J.\ Geom.\ Meth.\ Mod.\ Phys.\  {\bf 11} (2014) 1460006
[arXiv:1306.4426  ]; eConf C {\bf 0602061} (2006) 06
[Int.\ J.\ Geom.\ Meth.\ Mod.\ Phys.\  {\bf 4} (2007) 115]
[hep-th/0601213]; \\
V.~Faraoni and S.~Capozziello,
``Beyond Einstein gravity : A Survey of gravitational theories for cosmology
and astrophysics,''
Fundamental Theories of Physics, Vol. 170, Springer, 2010; \\
S.~Capozziello and M.~De Laurentis,
Phys.\ Rept.\  {\bf 509} (2011) 167
[arXiv:1108.6266  ]; \\
A.~de la Cruz-Dombriz and D.~Saez-Gomez,
Entropy {\bf 14} (2012) 1717
[arXiv:1207.2663  ].; \\
R.~Myrzakulov, L.~Sebastiani and S.~Zerbini,
  Int.\ J.\ Mod.\ Phys.\ D {\bf 22} (2013) 1330017
  [arXiv:1302.4646  ]

\bibitem{starobinsky} A.~A.~Starobinsky,  Phys.\ Lett.\ B {\bf 91} (1980)
99.;\\
J.~D.~Barrow and S.~Cotsakis,  Phys.\ Lett.\ B  214 (1988) 515.

\bibitem{encyclopedia}  J.~Martin, C.~Ringeval and V.~Vennin,
  Phys.\ Dark Univ.\  (2014)
  [arXiv:1303.3787 [astro-ph.CO]].



\bibitem{noh}  H.~Noh and J.~c.~Hwang,
  Phys.\ Lett.\ B {\bf 515} (2001) 231
  [astro-ph/0107069].; \\
  J.~c.~Hwang and H.~r.~Noh,
  Phys.\ Rev.\ D {\bf 65} (2002) 023512
  [astro-ph/0102005].; \\
  H.~Noh and J.~C.~Hwang,
  Mod.\ Phys.\ Lett.\ A {\bf 19} (2004) 1203.; \\
   J.~c.~Hwang and H.~Noh,
  Phys.\ Lett.\ B {\bf 506}, 13 (2001)
  [astro-ph/0102423].



\bibitem{sergeitraceanomaly}
A. Vilenkin, Phys. Rev. D { \bf 32}, 2511 (1985); \\
 K.~Bamba, R.~Myrzakulov, S.~D.~Odintsov and L.~Sebastiani,
  Phys.\ Rev.\ D {\bf 90} (2014) 4,  043505
  [arXiv:1403.6649 [hep-th]].



\bibitem{sergnoj} S. Nojiri, S. D. Odintsov, Phys.Rev. D68 (2003) 123512
[hep-th/0307288]



\bibitem{capp} S.~Capozziello, M.~De Laurentis and O.~Luongo,
  Int.\ J.\ Mod.\ Phys.\ D {\bf 24} (2014) 04,  1541002
  [arXiv:1411.2822 [gr-qc]]




\bibitem{planck}
P.~A.~R.~Ade {\it et al.}  [Planck Collaboration],
  arXiv:1502.02114 [astro-ph.CO]. ;\\
P.~A.~R.~Ade {\it et al.}  [Planck Collaboration],
Astron.\ Astrophys.\   571 (2014) A22 [arXiv:1303.5082
[astro-ph.CO]].

\bibitem{kaizer}

D.~I.~Kaiser,
  [astro-ph/9507048].;\\
   D.~I.~Kaiser,
  Phys.\ Rev.\ D {\bf 52}, 4295 (1995)
  [astro-ph/9408044].



\bibitem{Nojiri:2006gh}
S.~Nojiri and S.~D.~Odintsov,
Phys.\ Rev.\ D {\bf 74}, 086005 (2006)
[arXiv:hep-th/0608008].;J.\ Phys.\ Conf.\ Ser.\ {\bf 66}, 012005 (2007)
[arXiv:hep-th/0611071]


\bibitem{Capozziello:2006dj}
S.~Capozziello, S.~Nojiri, S.~D.~Odintsov and A.~Troisi,
Phys.\ Lett.\ B {\bf 639}, 135 (2006)
[arXiv:astro-ph/0604431].





\bibitem{sergbam08}
K.~Bamba, S.~Nojiri and S.~D.~Odintsov,
JCAP {\bf 0810} (2008) 045 [arXiv:0807.2575].




\bibitem{sergbam08bounce}

 K.~Bamba, A.~N.~Makarenko, A.~N.~Myagky, S.~Nojiri and S.~D.~Odintsov,
  JCAP {\bf 1401} (2014) 008
  [arXiv:1309.3748]



\bibitem{phantom} K. Bamba, Chao-Qiang Geng, S.Nojiri, S. D. Odintsov,
Phys.Rev.D79 (2009) 083014;\\
  Y. Du, H. Zhang, Xin-Zhou Li, Eur. Phys. J. C  71 (2011) 1660; \\
  S.~Nojiri and S.~D.~Odintsov,
  Gen.\ Rel.\ Grav.\  {\bf 38}, 1285 (2006)
  [hep-th/0506212].; \\
S.~Capozziello, S.~Nojiri and S.~D.~Odintsov,
Phys.\ Lett.\ B {\bf 632}, 597 (2006) [arXiv:hep-th/0507182].; \\
   Z.~G.~Liu and Y.~S.~Piao,
  Phys.\ Lett.\ B {\bf 713}, 53 (2012)
  [arXiv:1203.4901 [gr-qc]]. ;\\
  Y.~S.~Piao and Y.~Z.~Zhang,
  Phys.\ Rev.\ D {\bf 70} (2004) 063513
  [astro-ph/0401231].;\\
  Z.~K.~Guo, Y.~S.~Piao and Y.~Z.~Zhang,
  Phys.\ Lett.\ B {\bf 594} (2004) 247
  [astro-ph/0404225].;\\
  S.~Nesseris and L.~Perivolaropoulos,
  JCAP {\bf 0701}, 018 (2007)
  [astro-ph/0610092].



\bibitem{sergeinewplb}

W.~Hu and I.~Sawicki,
  Phys.\ Rev.\ D {\bf 76} (2007) 064004
  [arXiv:0705.1158 [astro-ph]]; \\
 S.~Nojiri and S.~D.~Odintsov,
  Phys.\ Lett.\ B {\bf 657} (2007) 238
  [arXiv:0707.1941 [hep-th]]



\bibitem{sergeiexp}

G.~Cognola, E.~Elizalde, S.~Nojiri, S.~D.~Odintsov, L.~Sebastiani and
S.~Zerbini,
  Phys.\ Rev.\ D {\bf 77} (2008) 046009
  [arXiv:0712.4017 [hep-th]].


\bibitem{LQC}
A.~Ashtekar and P.~Singh,
Class.\ Quant.\ Grav.\  {\bf 28} (2011) 213001
[arXiv:1108.0893 [gr-qc]]; \\
A.~Ashtekar,
Nuovo Cim.\ B {\bf 122} (2007) 135
[gr-qc/0702030]; \\
M.~Bojowald,
Class.\ Quant.\ Grav.\  {\bf 26} (2009) 075020
[arXiv:0811.4129 [gr-qc]]; \\
T.~Cailleteau, A.~Barrau, J.~Grain and F.~Vidotto,
Phys.\ Rev.\ D {\bf 86} (2012) 087301
[arXiv:1206.6736 [gr-qc]]; \\
J.~Quintin, Y.~F.~Cai and R.~H.~Brandenberger,
Phys.\ Rev.\ D {\bf 90} (2014) 6,  063507
[arXiv:1406.6049 [gr-qc]]; \\
Y.~F.~Cai, R.~Brandenberger and X.~Zhang,
Phys.\ Lett.\ B {\bf 703} (2011) 25
[arXiv:1105.4286 [hep-th]]; \\
Y.~F.~Cai, R.~Brandenberger and X.~Zhang,
JCAP {\bf 1103} (2011) 003
[arXiv:1101.0822 [hep-th]]; \\
K.~Bamba, J.~de Haro and S.~D.~Odintsov,
JCAP {\bf 1302} (2013) 008
[arXiv:1211.2968 [gr-qc]]; \\
J.~de Haro,
JCAP {\bf 1211} (2012) 037 [arXiv:1207.3621 [gr-qc]].

[arXiv:1108.0893 ]; A. Ashtekar, Nuovo Cim. B122 (2007) 135 [gr-qc/0702030];
M. Bojowald, Class.Quant.Grav. 26 (2009) 075020 [arXiv:0811.4129]; T.
Cailleteau, A. Barrau, J. Grain, F. Vidotto, Phys.Rev. D86 (2012) 087301
[arXiv:1206.6736]; J. Quintin, Yi-Fu Cai, R. H. Brandenberger, Phys. Rev.
D90 (2014) 063507 [arXiv:1406.6049] ; Yi-Fu Cai, R. Brandenberger, X. Zhang,
Phys.Lett. B703 (2011) 25 [arXiv:1105.4286] ; Yi-Fu Cai, R. Brandenberger,
X. Zhang, JCAP 1103 (2011) 003 [arXiv:1101.0822]







\end{thebibliography}
\end{document}